\documentclass[aps,prd,showkeys,twocolumn,superscriptaddress]{revtex4-2}
\usepackage{amsmath}
\usepackage{hyperref}
\usepackage{graphicx}
\usepackage{amsfonts}
\usepackage{changes}

\begin{document}


\title{Chaos and fast scrambling delays of dyonic Kerr-Sen-AdS$_4$ black hole and its ultra-spinning version}


\author{Hadyan Luthfan Prihadi}

\email{hadyanluthfanp@s.itb.ac.id}
\affiliation{Theoretical High Energy Physics Group, Department of Physics, Institut Teknologi Bandung, Jl. Ganesha 10 Bandung, Indonesia.}
\affiliation{Indonesia Center for Theoretical and Mathematical Physics (ICTMP), Institut Teknologi Bandung, Jl. Ganesha 10 Bandung,
	40132, Indonesia.}
	
\author{Freddy Permana Zen}
\email{fpzen@fi.itb.ac.id}
\affiliation{Theoretical High Energy Physics Group, Department of Physics, Institut Teknologi Bandung, Jl. Ganesha 10 Bandung, Indonesia.}
\affiliation{Indonesia Center for Theoretical and Mathematical Physics (ICTMP), Institut Teknologi Bandung, Jl. Ganesha 10 Bandung,
	40132, Indonesia.}
\author{Donny Dwiputra}
\email{donny.dwiputra@ymail.com}
\affiliation{Research Center for Quantum Physics, National Research and Innovation Agency (BRIN) South Tangerang 15314, Indonesia.}
\affiliation{Theoretical High Energy Physics Group, Department of Physics, Institut Teknologi Bandung, Jl. Ganesha 10 Bandung, Indonesia.}
\affiliation{Indonesia Center for Theoretical and Mathematical Physics (ICTMP), Institut Teknologi Bandung, Jl. Ganesha 10 Bandung,
	40132, Indonesia.}
\author{Seramika Ariwahjoedi}
\email{sera001@brin.go.id}
\affiliation{Asia Pacific Center for Theoretical Physics Pohang University of Science and Technology Pohang 37673, Gyeongsangbuk-do, South Korea}
\affiliation{Research Center for Quantum Physics, National Research and Innovation Agency (BRIN) South Tangerang 15314, Indonesia.}


\date{\today}

\begin{abstract}
The scrambling time and its delay are calculated using holography in an asymptotically AdS black hole solution of the gauged Einstein-Maxwell-Dilaton-Axion (EMDA) theory,  the dyonic Kerr-Sen-AdS$_4$ black hole, perturbed by rotating and charged shock waves along the equator. The leading term of the scrambling time for a black hole with large entropy is logarithmic in the entropy and hence supports the fast scrambling conjecture for this black hole solution, which implies that the system under consideration is chaotic. We also find that the instantaneous minimal Lyapunov index is bounded by $\kappa=2\pi T_H/(1-\mu\mathcal{L})$, which is analogous to the surface gravity but for the rotating shock waves, and becomes closer to equality for the near extremal black hole. For a small value of the AdS scale, we found that the Lyapunov exponent can exceed the bound for a large value of $\mathcal{L}$. Due to the presence of the electric and magnetic charge of the shock waves, we also show that the scrambling process of this holographic system is delayed by a time scale that depends on the charges of the shock waves. The calculations also hold for the ultra-spinning version of this black hole. The result of this paper generalizes the holographic calculations of chaotic systems which are described by an EMDA theory in the bulk.
\end{abstract}


\maketitle

\section{Introduction}
Due to their extreme properties, such as a strong gravitational field, black holes provide a good environment for studying the quantum effects/signatures of gravity. Previous studies of quantum gravity involve the study of string theory and loop quantum gravity \cite{Rovelli1990,Ariwahjoedi2015}. One of the other ways to investigate the quantum properties of gravity around a black hole is by understanding chaotic phenomena of holographic systems that are dual to some black holes, which has become an interesting topic until very recently \cite{Shenker2014, Leichenauer2014, Malvimat2022BTZ, Malvimat2022KerrAdS,Malvimat2022KerrAdS5,horowitz_bouncing_2022}. The study about chaotic phenomena of the 1-dimensional Sachdev-Ye-Kitaev (SYK) model \cite{MaldacenaStanford2016} which is conjectured to be dual to the 2-dimensional Jackiw-Teitelboim (JT) gravity as the effective theory of a near-extremal black hole near its horizon \cite{MaldacenaStanfordYang2016} also leads us to the understanding of a possibility to study traversable wormholes in the lab \cite{jafferis_traversable_2022,https://doi.org/10.48550/arxiv.2301.03522,PRXQuantum.4.010321}. According to the fast scrambling conjecture \cite{Sekino2008}, black holes are the fastest scrambler in nature, with the scrambling time proportional to the logarithm of its degrees of freedom or entropy, $t_*\sim\log S$, for systems with a large number of degrees of freedom (see also \cite{Hayden2007,https://doi.org/10.48550/arxiv.1101.6048,Lashkari2013}). A system can be considered chaotic when the Out-Of-Time-Ordered Correlators (OTOC) decay exponentially. The time required for the OTOC to vanish is known as the scrambling time, and for a fast scrambler, it is logarithmic in entropy.\\
\indent Another way to diagnose chaos in a quantum system is by using the Mutual Information $I(A;B)=S_A+S_B-S_{A\cup B}$, where $S_A$ is the von Neumann entropy for the reduced density matrix $\rho_A$ of a subsystem $A$. This can be done since $I(A;B)$ provides an upper bound for correlators. If $I(A;B) = 0$, then the correlators also vanish \cite{Wolf2008}, and the scrambling time can be obtained from the time required for $I(A;B)$ to vanish. Using the help from holographic theories such as the AdS/CFT correspondence \cite{Maldacena1997_LargeN}, the Mutual Information can be calculated using the holographic entanglement entropy, i.e. the Ryu-Takayanagi (RT) or Hubeny-Rangamani-Takayanagi (HRT) formula \cite{RyuTakayanagi1, RyuTakayanagi2,Hubeny_2007}. Entanglement entropy of a subregion in a conformal field theory (CFT) with an anti-de Sitter (AdS) dual is equal to the area of a minimal surface in the bulk which is homologous to the region divided by $4G_N$. This can be used to calculate the Mutual Information of a thermofield CFT which is dual to an eternal black hole in the AdS bulk \cite{MaldacenaEternal2003, Hartman2013}. Recent calculation of entanglement entropy using replica trick in the bulk shows that it is also possible to calculate entanglement entropy of black hole spacetimes with multiple horizons \cite{Prihadi_2023}.\\
\indent The chaotic behavior of a black hole is first studied in \cite{Shenker2014}, by calculating the Mutual Information in a CFT which is dual to a three-dimensional Banadoz-Teitelboim-Zanelli (BTZ) black hole (see also \cite{Shenker2014multipleshocks, Shenker2015}). A tiny perturbation traveling at the speed of light which is sent from the left boundary at a very early time and near the horizon can get highly blue-shifted and disrupt the geometry to make the Mutual Information vanish. Such a perturbation is called the gravitational shock waves which are represented by the Dray-'t Hooft solution \cite{Dray1985}. Adding the perturbation to the bulk geometry also corresponds to disrupting the thermofield double state in the boundary CFT, causing the correlator to vanish in late times. From \cite{Shenker2014}, it is found that the scrambling time for a black hole with large entropy is given by $t_*=\frac{\beta}{2\pi}\log\mathcal{S}$, where $\mathcal{S}$ is the black hole entropy and $\beta$ is the inverse Hawking temperature. Furthermore, the Lyapunov exponent is found to saturate the Maldacena, Shenker, and Stanford (MSS) bound on chaos \cite{Maldacena2016}, i.e. $\lambda_L=2\pi/\beta$. This observation supports the fast scrambling conjecture, which states that a black hole is maximally chaotic. The chaotic feature of black holes can be tested for other more general black holes such as charged and rotating black holes. The charged extension is first studied in \cite{Leichenauer2014}, where the scrambling time for a Reissner-Nordstr\"om-AdS black hole perturbed by neutral shock waves is calculated. It is also found that the scrambling time is logarithmic in entropy. The calculation is then generalized to charged and rotating BTZ black holes perturbed by charged shocks \cite{Reynolds2016}. Aside from the bulk calculations, the chaotic behavior of black holes has also been studied from the CFT side.\\
\indent A recent study \cite{horowitz_bouncing_2022} shows that the charged shell must bounce inside the horizon and change its null trajectory for the shock wave solution to not violate the null energy condition from the null-shell formalism \cite{Barrabes1991, Poisson2002}. The calculation is done for a four-dimensional Reissner-Nordstr\"om-AdS black hole perturbed by charged shock waves. This has an important implication for interpreting the scrambling time obtained. According to standard calculations, the effect of charged shock waves is to increase the scrambling time by an extra factor that depends on the charge of the shocks. However, since the shock waves bounce inside the horizon, there is a maximum time $t_b$ for a null particle sent from the right asymptotic to meet the shock waves inside the horizon. The difference between $t_b$ and the time the shocks are sent is equal to the difference between scrambling time with neutral shocks and charged ones plus some terms in order of the thermal time $\beta$. Using quantum circuit description, \cite{horowitz_bouncing_2022} concludes that the effect of the charged shock waves is to delay the start of the scrambling process.\\
\indent Other than the charged shock waves, one may also consider rotating shock waves. This is first done in \cite{Malvimat2022BTZ} for rotating BTZ black hole and in \cite{Malvimat2022KerrAdS, Malvimat2022KerrAdS5} for Kerr-AdS black hole in 4 and 5 dimensions. The importance of adding the angular momentum to the shocks is that the Lyapunov exponent can be larger than the MSS bound \cite{halder2019global} due to the existence of a global conserved charge. Furthermore, the solution survives the extremal limit even though the black hole temperature becomes zero at extremality.\\
\indent Aside from the standard Reissner-Nordstr\"om and Kerr black holes, there are many known black hole solutions that can be described microscopically using holographic theories \cite{Sakti2018, Sakti2019,Sakti2020, Prihadi2020,Sakti2021a, Sakti2021,Sakti2022KerrSen,Sakti2023}. It is very interesting to study the chaotic behavior of various black holes, particularly in an asymptotically AdS spacetime, to give us a better understanding of the chaotic properties of black holes or even holographic theories in general. It is shown in \cite{Malvimat2022KerrAdS} that the charged version (the Kerr-Newman-AdS) black hole does not provide much difference compared to the uncharged Kerr-AdS black hole. However, there is another black hole solution that is quite similar (to some extent) to the Kerr-Newman-AdS black hole, but with extra dilaton and axion scalar fields, named the dyonic Kerr-Sen-AdS$_4$ black hole. The chaotic behavior of such a black hole has not been studied before. Since the dilaton and axion fields are present, the result will differ from the charged and rotating Kerr-Newman-AdS black hole. Furthermore, it is widely known that rotating black holes in AdS have an ultra-spinning limit--a limit where the rotational parameter reaches its maximum value. It is also interesting to study the chaotic behavior of rotating black holes in such a limit since it is known \cite{Sakti2022KerrSen} that an ultra-spinning version of a dyonic Kerr-Sen-AdS black hole may still obey the Reverse Isoperimetric Inequality due to the existence of the dilaton and axion fields. Charged and rotating perturbations, in addition to the uncharged rotating perturbations, will also give a significant effect on the chaotic dynamics of the black holes, and it also has not been studied before, especially in a four-dimensional rotating and charged black holes with dilaton and axion fields.\\
\indent  In this work, we extend the calculations for the scrambling time for charged and rotating black holes in the gauged Einstein-Maxwell-Dilaton-Axion (EMDA) theory perturbed by (electrically and magnetically) charged rotating shock waves. The black hole solution to the gauged EMDA theory in 4 dimensions is called the dyonic Kerr-Sen-AdS$_4$ \cite{Wu2020}. The gauged version of the theory is equipped with a negative cosmological constant which provides us with an asymptotically AdS spacetime. Therefore, it is suitable for studying chaos in this black hole background using AdS/CFT. The eternal black hole form of this dyonic Kerr-Sen-AdS$_4$ black hole can also be represented holographically by two CFTs in their left and right asymptotic boundaries similar to \cite{MaldacenaEternal2003}, with extra chemical potentials which correspond to the rotation and (electric and magnetic) charges. Studying the chaotic behavior of such a system also lead us to an even more understanding of the holographic relation in the gauged EMDA theory. Using the Kerr/CFT correspondence \cite{Guica2009} and its extensions \cite{Sakti2018, Sakti2019,Sakti2020, Sakti2021}, the extremal dyonic Kerr-Sen-AdS$_4$ black hole entropy can be described microscopically by its dual two-dimensional CFT \cite{Sakti2022KerrSen,Sakti2023}. The dyonic Kerr-Sen-AdS$_4$ solution is parameterized by its mass, AdS scale, angular momentum, electric and magnetic charges, and two extra charges which correspond to the dilaton and axion fields.\\
\indent Using holographic calculations, we found that the scrambling time of the black hole is also logarithmic in entropy. The minimal instantaneous Lyapunov exponent $\lambda_L$ is bounded by $\kappa=2\pi T_H/(1-\mu\mathcal{L})$, where $T_H$ and $\mu$ are the temperature and the angular momentum of the black hole respectively, and $\mathcal{L}$ is the angular momentum (per unit energy) of the shock waves. This indicates that the system in the dyonic Kerr-Sen-AdS$_4$ black hole is chaotic and thus supports the fast scrambling conjecture, even for the rotating charged black hole in the EMDA theory. We show that the ratio $\kappa/\lambda_L$ is constant with respect to $\mathcal{L}$, and approaches 1 as the AdS radius $l$ becomes large. The constant can vary, depending on the black hole parameters, and thus we suggest that it should take the form $\kappa/\lambda_L=\mathcal{C}$, for some constant $\mathcal{C}>1$. For example, it has been shown recently that a three-dimensional rotating BTZ black hole has $\mathcal{C}=2$ \cite{Malvimat2022BTZ}. Interestingly, for a small value of $l$, $l=1$, we observe that the bound $\lambda_L<\kappa$ can be violated for large $\mathcal{L}$. This is unusual and it does not present in the case of an uncharged Kerr-AdS$_4$ black hole \cite{Malvimat2022KerrAdS}. The violations of chaos bound have been observed earlier (see, for example, \cite{Zhao2018, Gwak_2022,Yu2023Violating, Yu_2022}) using different methods, although only the standard MSS bound $2\pi T_H$ is used. The violations are present in the charged black hole cases, for some value of rotation parameter (both black hole's and particle's rotation). In this work, we show that the bound $\kappa/(1-\mu\mathcal{L})$ can also be violated due to the existence of the dilaton and axion charges.\\
\indent The structure of this paper is as follows: In Section \ref{section2}, we review the dyonic Kerr-Sen-AdS$_4$ black hole solution and its ultra-spinning version in \cite{Wu2020,Gnecchi2014} including their thermodynamics. In Section \ref{section3}, we construct the lightcone coordinates of the rotating and charged shockwaves and then obtain the corresponding Dray-'t Hooft shock waves solution for this geometry. In Section \ref{section4}, we first briefly explain the holographic CFT model for this dyonic Kerr-Sen-AdS$_4$ black hole. Using the RT/HRT surfaces, we calculate the corresponding entanglement entropy and obtain the Mutual Information which is sensitive to the charged and rotating shock waves in the bulk. From that, we can obtain the scrambling time $\tau_*$ and the scrambling delay $\tau_d$ both for the dyonic Kerr-Sen-AdS$_4$ black hole and its ultra-spinning counterpart. We also calculate the mutual information using holography, which leads to the calculations of the Lyapunov exponent. We show how the Lyapunov exponent behaves as we change the black hole parameters. These observations support the fast scrambling conjecture for the dyonic Kerr-Sen-AdS$_4$ black hole. We sum up our work in the Discussions and Conclusions session in Section \ref{section5}.
\section{Dyonic Kerr-Sen-AdS$_4$ metric}\label{section2}
In this section, we briefly review the black hole solution to the gauged Einstein-Maxwell-Dilaton-Axion theory known as the dyonic Kerr-Sen-AdS$_4$ black hole \cite{Wu2020}. We also write down the thermodynamic quantities and their first-law-like relation. The ultra-spinning version of the metric with $\varphi\rightarrow\varphi/\Xi$ and $a\rightarrow l$ is also given along with its thermodynamics. \\
\indent The dyonic Kerr-Sen-AdS$_4$ black hole is the solution to the gauged Einstein-Maxwell-Dilaton-Axion (EMDA) theory with action
\begin{widetext}
\begin{equation}
I=\frac{1}{16\pi G_N}\int\sqrt{|g|}d^4x\bigg(R-\frac{1}{2}(\partial\phi)^2-\frac{1}{2}e^{2\phi}(\partial\chi)^2-e^{-\phi}F^2+\frac{\chi}{2}F\tilde{F}\bigg)+I_\Lambda,
\end{equation}
\end{widetext}
where $\phi$ and $\chi$ are the dilaton scalar and axion pseudoscalar field respectively, $F=dA$ is the electromagnetic tensor of an abelian gauge potential one-form $A_\mu$ with its Hodge dual $\tilde{F}$. We use the notation $(\partial\phi)^2=(\partial_\mu\phi)(\partial^\mu\phi)$, $(\partial\chi)^2=(\partial_\mu\chi)(\partial^\mu\chi)$, $F^2=F_{\mu\nu}F^{\mu\nu}$, $F\tilde{F}=F_{\mu\nu}\tilde{F}^{\mu\nu}$ with $\tilde{F}^{\mu\nu}=\varepsilon^{\mu\nu\alpha\beta}F_{\alpha\beta}$ where $\varepsilon^{\mu\nu\alpha\beta}$ denotes a four-dimensional totally-antisymmetric tensor. The axion pseudoscalar $\chi$ is defined from $d\mathcal{B}=-e^{2\phi}\star d\chi$, where $\mathcal{B}_{\mu\nu}$ is an antisymmetric two-form tensor and $\star$ denotes the Hodge duality operator. The term $I_\Lambda$ represents the cosmological-constant term which comes from the gauged version of the standard EMDA theory and It is given by
\begin{equation}
I_\Lambda=\frac{1}{16\pi G_N}\int\sqrt{|g|}d^4x(4+e^{-\phi}+e^\phi(1+\chi^2))/l^2.
\end{equation}
\indent The solution for the dyonic Kerr-Sen-AdS$_4$ black hole is given by \cite{Wu2020}
\begin{equation}
ds^2=-\frac{\Delta}{\Sigma}X^2+\frac{\Sigma}{\Delta}dr^2+\frac{\Sigma}{\Delta_\theta}d\theta^2+\frac{\Delta_\theta\sin^2\theta}{\Sigma}Y^2,
\end{equation}
where
\begin{align}
X=&dt-\frac{a\sin^2\theta}{\Xi} d\varphi,\\\nonumber Y=&adt-\frac{(r^2-d^2-k^2+a^2)}{\Xi}d\varphi,\nonumber\\
\Delta(r)=&\bigg(1+\frac{r^2-d^2-k^2}{l^2}\bigg)(r^2-d^2-k^2+a^2)\\\nonumber&-2Mr+p^2+q^2,\nonumber\\
\Delta_\theta=&1-\frac{a^2}{l^2}\cos^2\theta,\;\;\;\Xi=1-\frac{a^2}{l^2},\\\nonumber
\Sigma=&r^2-d^2-k^2+a^2\cos^2\theta.
\end{align}
Here, $m,a,l,d,k,p,q$ are the mass, rotation, cosmological constant (AdS length), dilaton charge, axion charge, magnetic charge, and electric charge parameters respectively with the relations $d=(p^2-q^2)/2m$ and $k=pq/m$. Note that we use the shifted coordinate $r\rightarrow r+d$ for mathematical simplicity. The potential $A_\mu$, its dual $B_\mu$ defined from $e^{-\phi}\star F+\chi F=-dB$, dilaton scalar and axion pseudoscalar fields are given by
\begin{align}
A&=\frac{q(r+d-p^2/m)}{\Sigma}X-\frac{p\cos\theta}{\Sigma}Y,\nonumber\\
B&=\frac{p(r+d-p^2/m)}{\Sigma}X+\frac{q\cos\theta}{\Sigma}Y,\nonumber\\
e^\phi&=\frac{(r+d)^2+(a\cos\theta+k)^2}{\Sigma},\nonumber\\ \chi&=2\frac{kr-da\cos\theta}{(r+d)^2+(a\cos\theta+k)^2}.
\end{align}
\indent The thermodynamic quantities of the dyonic Kerr-Sen-AdS$_4$ black hole are given as follows:
\begin{align}\label{normalfunctions}
M&=\frac{m}{\Xi},\;\;\;J=\frac{ma}{\Xi},\;\;\;Q=\frac{q}{\Xi},\;\;\;P=\frac{p}{\Xi},\nonumber\\
S&=\frac{\pi}{\Xi}(r_+^2-d^2-k^2+a^2),\nonumber \\ \Omega_\varphi&=\frac{a\Xi}{r_+^2-d^2-k^2+a^2},\nonumber\\
\Phi&=\frac{q(r_++d-p^2/m)}{r_+^2-d^2-k^2+a^2},\nonumber \\ \Psi&=\frac{p(r_++d-p^2/m)}{r_+^2-d^2-k^2+a^2},\nonumber\\
2\pi T_H&=\frac{r_+(2r_+^2-2d^2-2k^2+a^2+l^2)-Ml^2}{(r_+^2-d^2-k^2+a^2)l^2},
\end{align}
where $r_+$ is the outermost horizon as the largest solution to $\Delta(r_+)=0$. Here, $M,J,Q,P$ are the mass, angular momentum, electric charge, and magnetic charge of the black holes as the corresponding conserved charges. Furthermore, $S$ is the entropy of the black hole while $\Omega_\varphi,\Phi,\Psi$ denotes the angular momentum, electric potential, and magnetic potential of the horizon respectively. For now, we assume that the black hole is non-extremal, i.e. $\Delta'(r_+)\neq0$ where the prime denotes the derivative with respect to $r$. In this work, we treat the cosmological constant parameter $l$ as non-dynamical such that the first law equation becomes
\begin{equation}
dM=T_H dS+\Omega_\varphi dJ+\Phi dQ+\Psi dP+Jd\Xi/(2a).
\end{equation}
The dyonic Kerr-Sen-AdS$_4$ metric is rotating at $r\rightarrow\infty$ with angular velocity
\begin{equation}
\Omega_\infty=-\frac{a}{l^2}.
\end{equation}
Therefore, a stationary observer at infinity can be obtained by a coordinate transformation $\varphi\rightarrow\varphi+\Omega_\infty t$ and some of the thermodynamic quantities are shifted by
\begin{equation}
M\rightarrow \bar{M}=M+\frac{a}{l^2},J=\frac{m}{\Xi^2},\;\;\;\Omega_\varphi\rightarrow\Omega_\varphi-\Omega_\infty,
\end{equation}
where now the newly shifted black hole angular velocity is defined as
\begin{equation}
\mu\equiv\Omega_\varphi-\Omega_\infty=\frac{a(1+(r_+^2-d^2-k^2)/l^2)}{r_+^2-d^2-k^2+a^2}.\label{mu}
\end{equation}
Now the angular velocity $\mu$ becomes the new chemical potential for the first-law relation
\begin{equation}
d\bar{M}=T_HdS+\mu dJ+\Phi dQ+\Psi dP.\label{firstlaw}
\end{equation}
\subsection{Ultra-spinning dyonic Kerr-Sen-AdS$_4$}
The redefinition of the coordinate $\varphi\rightarrow\varphi/\Xi$ allows us to obtain the ultra-spinning limit $a\rightarrow l$ of the dyonic Kerr-Sen-AdS$_4$ black hole, which now has the metric and the corresponding fields in the form of
\begin{align}\label{uspinmetric}
d\hat{s}^2&=-\frac{\hat{\Delta}}{\hat{\Sigma}}\hat{X}^2+\frac{\hat{\Sigma}}{\hat{\Delta}}dr^2+\frac{\hat{\Sigma}}{\sin^2\theta}d\theta^2+\frac{\sin^4\theta}{\hat{\Sigma}}\hat{Y}^2,\nonumber\\
A&=\frac{q(r+d-p^2/m)}{\hat{\Sigma}}\hat{X}=\frac{p\cos\theta}{\hat{\Sigma}}\hat{Y},\nonumber\\
B&=\frac{p(r+d-p^2/m)}{\hat{\Sigma}}\hat{X}+\frac{q\cos\theta}{\hat{\Sigma}}\hat{Y},\nonumber\\
e^\phi&=\frac{(r+d)^2+(l\cos\theta+k)^2}{\hat{\Sigma}},\\\nonumber\chi&=2\frac{kr-dl\cos\theta}{r^2+(l\cos\theta+k)^2},
\end{align}
with
\begin{align}
\hat{X}&=dt-l\sin^2\theta d\varphi,\\\nonumber\hat{Y}&=ldt-(r^2-d^2-k^2+l^2)d\varphi,\nonumber\\
\hat{\Delta}(r)&=(r^2-d^2-k^2+l^2)^2/l^2-2Mr+p^2+q^2,\nonumber\\
\hat{\Sigma}&=r^2-d^2-k^2+l^2\cos^2\theta.\label{uspinmetric2}
\end{align}
The periodicity of $\varphi$ is now assumed to be given by $\lambda$ instead of $2\pi$. Notice that the standard dyonic Kerr-Sen-AdS$_4$ metric $ds^2$ cannot be obtained directly from the ultra-spinning metric $d\hat{s}^2$.\\
\indent The thermodynamic quantities of the ultra-spinning dyonic Kerr-Sen-AdS$_4$ black hole are given by
\begin{align}
\hat{M}&=\frac{\mu}{2\pi}m,\;\;\hat{J}=\frac{\mu}{2\pi}ml=Ml,\;\;\hat{Q}=\frac{\mu}{2\pi}q,\;\;\hat{P}=\frac{\mu}{2\pi}p,\nonumber\\
\hat{S}&=\frac{\mu}{2}(r_+^2-d^2-k^2+l^2),\;\;\hat{\Omega}_\varphi=\frac{l}{r_+^2-d^2-k^2+l^2},\nonumber\\
\hat{\Phi}&=\frac{q(r_++d-p^2/m)}{r_+^2-d^2-k^2+l^2},\;\;\;\hat{\Psi}=\frac{p(r_++d-p^2/2m)}{r_+^2-d^2-k^2+l^2},\nonumber\\
2\pi\hat{T}_H&=2\frac{r_+}{l^2}-\frac{m}{r_+^2-d^2-k^2+l^2}.\label{ultraspinningfunction}
\end{align}
Again, $r_+$ is defined as the outermost horizon satisfying $\hat{\Delta}(\hat{r}_+)=0$ and we assume that the ultra-spinning black hole is also non-extremal with $\hat{\Delta}'(r_+)\neq0$. The first-law-like relation is now given by
\begin{equation}
d\hat{M}=\hat{T}_Hd\hat{S}+\hat{\Omega}_\varphi d\hat{J}+\hat{\Phi}d\hat{Q}+\hat{\Psi}d\hat{P}+\hat{K}d\lambda,\label{firstlawultraspinning}
\end{equation}
where $\lambda$ also plays a role as the dynamical variable with a new chemical potential
\begin{equation}
K=m\frac{l^2-(r_++d+q^2/m)(r_++d-p^2/m)}{4\pi(r_+^2-d^2-k^2+l^2)}.
\end{equation}
\section{Equatorial Rotating Shock wave Solutions}\label{section3}
\subsection{Kruskal coordinates}
Consider an equatorial ($\theta=\pi/2$) rotating null particle in the background of the dyonic Kerr-Sen-AdS$_4$ black hole with unit energy $\mathcal{E}=1$ and angular momentum per unit energy $\mathcal{L}$. We will then call this the rotating shock wave. Such a particle will have a geodesic $\xi^\mu$ observed by a stationary observer at infinity which satisfies
\begin{equation}
\xi^2=0,\;\;\;\xi\cdot\zeta_t=-\mathcal{E},\;\;\;\xi\cdot\zeta_\varphi=\mathcal{L},\;\;\;\xi^\theta=0.\label{geodesicconstraint}
\end{equation}
The last equality is required for the equatorial orbit, which means that the particle always stays at the equator $\theta=\pi/2$. The corresponding Killing vector associated with axisymmetry is given by $\zeta_t=\partial_t-a/l^2\partial_\varphi$ and $\zeta_\varphi=\partial_\varphi$. There are two solutions to eq. (\ref{geodesicconstraint}), $\xi_\pm^\mu$, where the negative solution is obtained from reversing the axisymmetry $-\mathcal{E}\rightarrow\mathcal{E}$ and $\mathcal{L}\rightarrow-\mathcal{L}$. The geodesic solution in the background of a dyonic Kerr-Sen-AdS$_4$ black hole for unit energy ($\mathcal{E}=1$) is given by
\begin{equation}
\xi_+\cdot dx=dr_*-d\tau\equiv du,\;\;\;\xi_-\cdot dx=dr_*+d\tau\equiv dv,\label{uvcoordinates}
\end{equation}
where $\tau=(1-a\mathcal{L}/l^2)t-\mathcal{L}\varphi$ and $r_*$ is the tortoise-like coordinate given by
\begin{equation}
r_*(r)=\int\frac{\tilde{f}}{\Delta}dr,\label{tortoise}
\end{equation}
with $\tilde{f}$ is defined from
\begin{widetext}
\begin{equation}
\tilde{f}^2=-\Delta(\mathcal{L}-a)^2+[\mathcal{L}a(1+(r^2-d^2-k^2)/l^2)-(r^2-d^2-k^2+a^2)]^2.
\end{equation}
\end{widetext}
With vanishing dilaton and axion charge, i.e. $d,k\rightarrow0$, we recover the tortoise-like coordinate found in \cite{Malvimat2022KerrAdS} and with vanishing angular momentum $\mathcal{L}\rightarrow0$, we recover the standard Kruskal coordinates for rotating dyonic Kerr-Sen-AdS$_4$ black hole.\\
\indent The Kruskal-like metric which follows the geodesic of the rotating shock wave is given by
\begin{equation}
ds^2=F(r)dudv+\tilde{h}(r)(d\varphi+\tilde{h}_\tau(r)d\tau)^2,
\end{equation}
with
\begin{widetext}
\begin{align}
F(r)&=\frac{\Delta(r^2-d^2-k^2)}{\tilde{f}^2},\;\;\;\tilde{h}(r)=\frac{(1-a\mathcal{L}/l^2)^{-2}}{r^2-d^2-k^2}\frac{\tilde{f}^2}{\Xi^2},\nonumber\\
\tilde{h}_\tau(r)&=\frac{\Xi}{\tilde{f}^2}\big((\Delta(a-\mathcal{L})+a\big(a\mathcal{L}(1+(r^2-d^2-k^2)/l^2)-(r^2-d^2-k^2+a^2)\big)\big).
\end{align}
\end{widetext}
We then shift the axial coordinate $\varphi\rightarrow\eta z+\gamma\tau$ for two reasons: one is to make sure that $\tilde{h}_\tau(r)+\gamma$ behaves like $\mathcal{O}(r-r_+)$ near the horizon and to recover the black hole's horizon area when integrating $\theta$ from $0$ to $\pi$ and $z$ from $0$ to $2\pi$. Thus, we have
\begin{equation}
\gamma=\frac{\Omega_\varphi}{1-\mu\mathcal{L}},\;\;\;\eta=\frac{1}{1-\mu\mathcal{L}}.
\end{equation}
From the coordinate transformations, we have
\begin{equation}
ds^2=F(r)dudv+h(r)(dz+h_\tau(r)d\tau)^2,
\end{equation}
where $h(r)=\eta^2\tilde{h}(r)$ and $h_\tau(r)\equiv\eta^{-1}(\tilde{h}_\tau(r)+\gamma)$.\\
\indent Furthermore, we would like to work with coordinates that are affine at the horizon to generate the Dray-'t Hooft solution later on. In this case, $\{u,v\}$ coordinate is not affine, i.e.  the tangent vector $\chi_u=\partial_u$ satisfy $\chi_u\cdot\nabla\chi_u^\mu=\mathcal{K}\chi_u^\mu$ (similar with $\chi_v^\mu$), where
\begin{equation}
\mathcal{K}=\frac{1}{2}\xi_\pm\cdot\partial F.
\end{equation}
The new coordinates $\{U,V\}$ are
\begin{equation}
U=-e^{\kappa u},\;\;\;V=e^{\kappa v},\;\;\;\kappa\equiv\mathcal{K}|_{r=r_+},\label{affinecoord}
\end{equation}
and it can easily be shown that both $\partial_U$ and $\partial_V$ are affine coordinates at $r_+$. The metric in the affine coordinates is now written as
\begin{equation}
ds^2=\frac{F(r)}{\kappa^2UV}dUdV+h(r)(dz+h_\tau(r)d\tau)^2.\label{kruskal}
\end{equation}
The value of $\kappa$ is related to the black hole's surface gravity $\kappa_0$ by
\begin{equation}
\kappa=\frac{\kappa_0}{1-\mu\mathcal{L}}=\frac{2\pi T_H}{1-\mu\mathcal{L}}.
\end{equation}
This $\kappa$ has a form similar to the standard Kerr-AdS$_4$ black hole with rotating shock waves. However, it differs in the value of the Hawking temperature $T_H$ and the angular momentum $\mu$ which now depends on the value of the dilaton and axion charge $d,k$.\\
\indent Next, we would like to generate the Kruskal-like coordinates for the ultra-spinning counterpart where the metric is given by eq. (\ref{uspinmetric}). Since the ultra-spinning version of the dyonic Kerr-Sen-AdS$_4$ has a null $\varphi$ coordinate at the boundary $r\rightarrow\infty$ and hence the metric is non-diagonal, we need to adopt the so-called conformal completion technique to compute the conserved quantities \cite{Chen2006, Wu2020}. The mass and the angular momentum of the ultra-spinning dyonic Kerr-Sen-AdS$_4$ black hole given by eq. (\ref{ultraspinningfunction}) is obtained by calculating the conserved charge $\mathcal{Q}[\zeta]$ defined in eq. (9) of \cite{Wu2020} for a Killing vector $\zeta$. The timelike Killing vector $\zeta_t=\partial_t$ gives us the mass $\hat{M}$, while the axial Killing vector $\zeta_\varphi=\partial_\varphi$ gives us the angular momentum $\hat{J}$. Therefore, we also have to use the Killing vectors at infinity $\zeta_t=\partial_t$ and $\zeta_\varphi=\partial_\varphi$ which corresponds to the conserved energy $\mathcal{E}$ and angular momentum $\mathcal{L}$ of a rotating shock wave around the ultra-spinning black hole. 
\\
\indent The geodesic of a rotating shock wave $\hat{\xi}^\mu$ in the background of an ultra-spinning dyonic Kerr-Sen-Black hole is obtained by solving eq. (\ref{geodesicconstraint}) with the ultra-spinning black hole metric in eqs (\ref{uspinmetric}) and (\ref{uspinmetric2}) and Killing vectors $\zeta_t=\partial_t,\zeta_\varphi=\partial_\varphi$. The solutions for unit energy $\mathcal{E}=1$ and an angular momentum per unit energy $\mathcal{L}$ are given in the form of
\begin{equation}
\hat{\xi}_+\cdot dx=d\hat{r}_*-d\hat{\tau}\equiv d\hat{u},\;\;\;\hat{\xi}_-\cdot dx=d\hat{r}_*+d\hat{\tau}\equiv d\hat{v},
\end{equation}
where $\hat{\tau}=t-\mathcal{L}\varphi$ and
\begin{equation}
r_*\equiv\int\frac{\hat{\tilde{f}}dr}{\hat{\Delta}},
\end{equation}
with
\begin{equation}
\hat{\tilde{f}}^2=-\hat{\Delta}(\mathcal{L}-l)^2+\big(\mathcal{L}l-(r^2-d^2-k^2+l^2)\big)^2.
\end{equation}
Notice that both $\hat{\tau}$ and $\hat{\tilde{f}}$ cannot be obtained by taking $a\rightarrow l$ limit of $\tilde{f}$. The Kruskal-like metric for the ultra-spinning black hole with rotating shock waves is given by
\begin{equation}
d\hat{s}^2=\hat{F}(r)d\hat{u}d\hat{v}+\hat{\tilde{h}}(r)(d\varphi+\hat{\tilde{h}}_{\hat{\tau}}(r)d\hat{\tau})^2,
\end{equation}
where now we have
\begin{align}
\hat{F}(r)&=\frac{\hat{\Delta}(r^2-d^2-k^2)}{\hat{\tilde{f}}^2},\;\;\;\hat{\tilde{h}}(r)=\frac{\tilde{f}^2}{r^2-d^2-k^2},\nonumber\\
\hat{\tilde{h}}_{\hat{\tau}}(r)&=\frac{-\hat{\Delta}(\mathcal{L}-l)^2+l(\mathcal{L}l-(r^2-d^2-k^2+l^2))}{\tilde{f}^2}.
\end{align}
Following similar reasoning to the normal dyonic Kerr-Sen-AdS$_4$, we redefine the coordinate $\varphi\rightarrow\hat{\eta}\hat{z}+\hat{\gamma}\hat{\tau}$, where now
\begin{equation}
\hat{\eta}=\frac{1}{1-\hat{\Omega}_\varphi\mathcal{L}},\;\;\;\hat{\gamma}=\frac{\hat{\Omega}_\varphi}{1-\hat{\Omega}_\varphi\mathcal{L}}.\label{etaultraspinning}
\end{equation}
Again, all of the functions appear in the ultra-spinning case $\{\hat{F},\hat{\tilde{h}},\hat{\tilde{h}}_{\hat{\tau}},\hat{\eta},\hat{\gamma}\}$ cannot be directly obtained from $\{F,\tilde{h},\tilde{h}_\tau,\eta,\gamma\}$ by taking the limit of $a\rightarrow l$ and thus they are all completely different functions. After the coordinate transformation $\varphi\rightarrow\hat{\eta}\hat{z}+\hat{\gamma}\hat{\tau}$, the Kruskal-like metric for the ultra-spinning black hole now takes the form
\begin{equation}
d\hat{s}^2=\hat{F}(r)d\hat{u}d\hat{v}+\hat{h}(r)(d\hat{z}+\hat{h}_{\hat{\tau}}(r)d\hat{\tau})^2,
\end{equation}
where $\hat{h}(r)=\hat{\eta}^2\hat{\tilde{h}}(r)$ and $\hat{h}_{\hat{\tau}}(r)=\hat{\eta}^{-1}(\hat{\tilde{h}}_{\hat{\tau}}+\hat{\gamma})$.
\\
\indent We also would like to work with affine coordinates $\{\hat{U},\hat{V}\}$ instead of $\{\hat{u},\hat{v}\}$ for the ultra-spinning black hole. Following similar reasoning as before, we obtain
\begin{equation}
\hat{U}=-e^{\hat{\kappa}\hat{u}},\;\;\;\hat{V}=e^{\hat{\kappa}\hat{v}},
\end{equation}
where now $\hat{\kappa}$ is given by
\begin{equation}
\hat{\kappa}=\frac{2\pi\hat{T}_H}{(1-\hat{\Omega}_\varphi\mathcal{L})}.
\end{equation}
The metric in the affine coordinates $\{\hat{U},\hat{V}\}$ is now given by
\begin{equation}
d\hat{s}^2=\frac{\hat{F}(r)}{\hat{\kappa}^2\hat{U}\hat{V}}d\hat{U}d\hat{V}+\hat{h}(r)(d\hat{z}+\hat{h}_{\hat{\tau}}(r)d\hat{\tau})^2.
\end{equation}
In this coordinate system, we then generate the Dray-'t Hooft solution for the ultra-spinning dyonic Kerr-Sen-AdS$_4$ black hole.\\
\indent After we obtain the metric in Kruskal coordinates for the dyonic Kerr-Sen-AdS$_4$ and its ultra-spinning counterpart, we now see how the functions $F(r)$ and $\hat{F}(r)$ behave near the horizon. Indeed, by expanding $r$ near $r_+$ up to the second order, we have
\begin{align}
F(r)=&F'(r_+)(r-r_+)+\\\nonumber&\frac{1}{2}F''(r_+)(r-r_+)^2+...\;,\\
\hat{F}(r)=&\hat{F}'(r_+)(r-r_+)+\\\nonumber&\frac{1}{2}\hat{F}''(r_+)(r-r_+)^2+...\;,
\end{align}
with all $F'(r_+),F''(r_+),\hat{F}'(r_+),\hat{F}''(r_+)$ are non zero. Using the definitions of the Kruskal coordinates in eq. (\ref{affinecoord}) and the tortoise-like coordinates in (\ref{tortoise}) for the dyonic Kerr-Sen-AdS$_4$ black hole, we may write that, near the horizon,
\begin{align}
(r-r_+)F'(r_+)&=-\mathbb{A}UV,\label{Fprimed}
\end{align}
where $\mathbb{A}$ is some dimensionless proportionality constant which depends on the dyonic Kerr-Sen-AdS$_4$ black hole parameters. \\
\indent One need to be more cautious when considering $\hat{U}\hat{V}$ near the horizon for the ultra-spinning black hole because $\hat{\kappa}$ is now equipped with $\hat{\Omega}_\varphi$ instead of $\mu$. However, it turns out that the value of $\hat{U}\hat{V}$ near the horizon is still linear in $r-r_+$ since the tortoise coordinate near the horizon behaves as
\begin{align}
r_*&\approx\frac{(1-\hat{\Omega}_\varphi\mathcal{L})(r_+^2-d^2-k^2)}{\hat{\Delta}'(r_+)}\ln|r-r_+|+C\\\nonumber&=\frac{1}{2\hat{\kappa}}\ln|r-r_+|+C,
\end{align}
where $C$ is an integration constant in which the explicit value depends on the geometry of the black hole. Therefore, $\hat{U}\hat{V}\sim\mathcal{O}(r-r_+)$ near the horizon and we may also write
\begin{equation}
(r-r_+)\hat{F}'(r_+)=-\hat{\mathbb{A}}\hat{U}\hat{V},\label{Fprimedhat}
\end{equation}
where $\hat{\mathbb{A}}$ is a dimensionless proportionality constant that depends on the ultra-spinning black hole parameters.\\
\indent The expansions of the functions $F(r)$ and $\hat{F}(r)$ near the horizon can then be extended to the second order of $r-r_+$. For the standard dyonic Kerr-Sen-AdS$_4$, we have
\begin{equation}
\frac{F}{\mathbb{A}UV}=-1-\frac{\mathbb{A}UV}{2}\frac{F''(r_+)}{F'(r_+)}+\mathcal{O}(U^2V^2),
\end{equation}
while for the ultra-spinning black hole, we have
\begin{equation}
\frac{\hat{F}}{\hat{\mathbb{A}}\hat{U}\hat{V}}=-1-\frac{\hat{\mathbb{A}}\hat{U}\hat{V}}{2}\frac{\hat{F}''(r_+)}{\hat{F}'(r_+)}+\mathcal{O}(\hat{U}^2\hat{V}^2).
\end{equation}
\subsection{Extremal limits}
From the turning-point analysis, the maximum value that $\mathcal{L}$ can achieve if we need the perturbation to reach the horizon from infinity is bounded by $\mu^{-1}$ ($\mathcal{L}_{\text{max}}\leq\mu^{-1}$) and the equality holds at extremality, i.e. $\mathcal{L}_{\text{max}}=\mu^{-1}$ for $T_H\rightarrow0$ \cite{Malvimat2022KerrAdS}. For the dyonic Kerr-Sen-AdS$_4$ black hole, the value of $\kappa$ also survives the extremal limit.  By expanding the value of $r_+$ near the extremal horizon radius $r_0$, where $\Delta(r_0)=\Delta'(r_0)=0$, to the first-order approximation, we have
\begin{equation}
T_H\approx\frac{\partial T_H}{\partial r_+}\bigg|_{r_0}(r_+-r_0).
\end{equation}
On the other hand, the value of $(1-\mu\mathcal{L})$ can also be expanded near the extremal value $r_0$,
\begin{equation}
1-\mu\mathcal{L}\approx\bigg(\frac{1}{\mu}\frac{\partial\mu}{\partial r_+}\bigg)_{r_0}(r_+-r_0).
\end{equation}
Therefore, by taking the extremal limit $r_+\rightarrow r_0$, we obtain the extremal value of $\kappa$, 
\begin{equation}
\kappa_{\text{ext}}=\frac{2\pi (\partial T_H/\partial r_+)_{r_0}}{(\mu^{-1}\partial\mu/\partial r_+)_{r_0}}=-2\pi T_L\mu_{\text{ext}},
\end{equation}
where $\mu_{\text{ext}}$ is given by eq. (\ref{mu}) with $r_+$ is replaced by $r_0$ and $\kappa_{\text{ext}}$ is related by the definition of the left-moving Frolov-Thorne temperature found by the Kerr/CFT correspondence for the extremal dyonic Kerr-Sen-AdS$_4$ black hole $T_L\equiv-\frac{(\partial T_H/\partial r_+)_{r_0}}{(\partial\mu/\partial r_+)_{r_0}}$ derived in \cite{Sakti2022KerrSen, Sakti2023}. \\
\indent For the dyonic Kerr-Sen-AdS$_4$ black hole, we have
\begin{widetext}
\begin{equation}
\kappa_{\text{ext}}=\frac{(1+(r_0^2-d^2-k^2)/l^2)(1+a^2/l^2+(6r_0^2-2d^2-2k^2)/l^2)}{2r_0\Xi},
\end{equation}
\end{widetext}
while for the ultra-spinning counterpart (following the same procedure), we have
\begin{align}
\hat{\kappa}_{\text{ext}}&=-2\pi\hat{T}_L\hat{\Omega}^{\text{ext}}_\varphi\\&=\frac{(6r_+^2-2d^2-2k^2+2l^2)}{2r_+l^2}.
\end{align}
Both $\kappa_{\text{ext}}$ and $\hat{\kappa}_{\text{ext}}$ scales with the left-moving Frolov-Thorne temperature calculated by the Kerr/CFT. The difference lies in the type of angular momentum which appear in both $\kappa_{\text{ext}}$ and $\hat{\kappa}_{\text{ext}}$ where it is given by $\mu_{\text{ext}}$ for the former and $\hat{\Omega}^{\text{ext}}_\varphi$ for the latter (instead of $\hat{\mu}_{\text{ext}}$). This difference lies in the definition of angular velocities. For the standard black hole, $\mu$ is the difference between the horizon's angular velocity and the angular velocity of a stationary observer at infinity. For the ultra-spinning black hole, however, we do not have a notion of the angular velocity of a stationary observer at infinity because the $\varphi$ coordinate becomes null. Instead, what appears in $1-\hat{\Omega}_\varphi\mathcal{L}$ is the horizon's angular velocity $\hat{\Omega}_\varphi$ which is the chemical potential for $\hat{J}$. Up until this point, all of the hatted functions for the ultra-spinning black hole do not depend on the new dynamical variable $\lambda$.
\subsection{Dray-'t Hooft solution}
After we obtain the Kruskal-like coordinates for rotating null geodesic for both standard dyonic Kerr-Sen-AdS$_4$ black hole and its ultra-spinning counterpart, we are now able to generate the Dray-'t Hooft solution. For a Kruskal-like metric perturbed by an in-falling rotating shock wave at the equator located at $U_0$ for the standard black hole and at $\hat{U}_0$ for the ultra-spinning black hole, the metric is given by the shift for $U>U_0$,
\begin{equation}
V\rightarrow\tilde{V}=V+\alpha\Theta(U-U_0),
\end{equation}
for the dyonic Kerr-Sen-AdS$_4$, and 
\begin{equation}
\hat{V}\rightarrow\hat{\tilde{V}}=\hat{V}+\hat{\alpha}\Theta(\hat{U}-\hat{U}_0),\label{Vshift}
\end{equation}
for its ultra-spinning version. It is known that the shift should contain a function $f(\theta)$ which captures the perturbation away from the equator. However, when we only consider $\theta=\pi/2$, we may normalize the function such that $f(\pi/2)=1$. In this section, we will calculate the Dray-'t Hooft solution and the strength of the shock waves $\alpha$ for the standard dyonic Kerr-Sen-AdS$_4$ first, then followed by its ultra-spinning counterpart later on.
\subsubsection*{Standard Kerr-Sen-AdS$_4$}
We first present the coordinate $\tau$ in terms of $U$ and $V$. The metric in eq. (\ref{kruskal}) can then be written as
\begin{align}\label{metricUV}
ds^2=&\frac{F(r)}{\kappa^2UV}dUdV+\\\nonumber&h(r)\bigg(dz+\frac{h_\tau(r)}{2\kappa UV}(UdV-VdU)\bigg)^2.
\end{align}
After applying the Dray-'t Hooft solution which gives the shift in eq. (\ref{Vshift}) due to the shock, the metric becomes
\begin{equation}
ds^2\rightarrow\tilde{ds}^2-\frac{F}{\kappa^2UV}\delta(U)dU^2,
\end{equation}
after the shift, where $\tilde{ds}^2$ is given by eq. (\ref{kruskal}) with $V$ is replaced by $\tilde{V}$.\\
\indent The strength of the shift, $\alpha$ and $\hat{\alpha}$ can be obtained by requiring that the transverse volume element $H=\sqrt{g_{\theta\theta}g_{zz}}$ (in the full coordinates involving $\theta$) to be smooth at the location of the shock. Suppose that after the shock wave enters the black hole horizon, it increases the black hole mass by a small amount $E_0$. To make the case more general, we may also consider a perturbation which increases the value of the electric charge of the black hole by a small amount $\delta q$ and the magnetic charge by a small amount $\delta p$ as well although we assume that the trajectory of the perturbation still follows the rotating shock waves geodesic given by eq. (\ref{uvcoordinates}). In this case, the area of the horizon changes according to the first law of thermodynamics, which also change $H$ at the horizon, since $H$ is nothing but the area of the horizon, divided $4\pi$. By imposing $H$ to be smooth at the horizon, we have
\begin{equation}
H\big|_{U_0^+}=H\big|_{U_0-}.
\end{equation}
Using eq. (\ref{Fprimed}) and taking the limit $E_0\rightarrow0$ and $U_0\rightarrow0$ simultaneously, we have
\begin{equation}
\alpha=\frac{(\tilde{H}_{r_+}-H_{r_+})F'(r_+)}{U_0\mathbb{A}\tilde{H}'(r_+)},\label{alpharaw}
\end{equation}
where $\tilde{H}_{r_+}$ is the shifted horizon area divided by $4\pi$. Upon taking the limit of $E_0\rightarrow0$, we have $(\tilde{H}_{r_+}-H_{r_+})\rightarrow0$ as well. However, $\alpha$ can be fixed by taking the limit $U_0\rightarrow0$, which means that the shocks hover very close to the horizon, and were sent from the AdS boundary in the far past $\tau_0\rightarrow\infty$.\\
\indent To be more precise, since $H_{r_+}$ is proportional to the area of the horizon, we have
\begin{equation}
\tilde{H}_{r_+}-H_{r_+}\propto\delta S=\beta_H(\delta \bar{M}-\mu\delta J-\Phi\delta Q-\Psi\delta P),
\end{equation}
by the first law of thermodynamics given by eq. (\ref{firstlaw}), where $\beta_H=T_H^{-1}$ is the inverse of the Hawking temperature. The change in mass $\delta\bar{M}$ is proportional to the energy of the shock wave $E_0$ and the change in the angular momentum $\delta J=\delta\bar{M}\mathcal{L}$ is proportional to $E_0\mathcal{L}$. From the thermodynamic relation between $Q,P$ and $q,p$, we have $\delta Q\sim\delta q$ and $\delta P\sim\delta p$ which is small, i.e. $\delta q,\delta p\rightarrow0$, but can be fixed by setting $\mathcal Q=\delta q/E_0$ and $\mathcal{P}=\delta p/E_0$ with $E_0\rightarrow0$ as well. Therefore, we may write
\begin{equation}
\tilde{H}_{r_+}-H_{r_+}\sim\beta_H\delta\bar{M}(1-\mu\mathcal{L}-\Phi\mathcal{Q}-\Psi\mathcal{P}),
\end{equation}
where now, aside from the angular momentum per unit energy $\mathcal{L}$, we also have the electric and magnetic charge per unit energy, $\mathcal{Q},\mathcal{P}$. Next, we need to calculate $F'(r_+)/\tilde{H}'(r_+)$. It is straightforward to show that, for large black hole entropy $S\rightarrow\infty$, it is given by
\begin{equation}
\frac{F'(r_+)}{\tilde{H}'(r_+)}=\frac{\mathbb{B}}{S},
\end{equation}
where $\mathbb{B}$ is some function of the black hole parameters and it is dimensionless.\\
\indent From the thermodynamic analysis, we obtain (by absorbing all of the proportionality constants $\mathbb{A},\mathbb{B},2\pi$ into the definition of $\delta\bar{M}$)
\begin{equation}
\alpha=\frac{\beta_H E_0(1-\mu\mathcal{L}-\Phi\mathcal{Q}-\Psi \mathcal{P})}{U_0S},\label{alpha}
\end{equation}
which is set to be fixed upon taking $E_0\rightarrow0$ and $U_0\rightarrow0$ for large $S$.\\\\
\subsubsection*{Ultra-spinning dyonic Kerr-Sen-AdS$_4$}
We may follow similar reasoning to calculate $\alpha$ for the ultra-spinning case. The value of $\hat{\alpha}$ is still given by the form of eq. (\ref{alpharaw}), with all functions hatted, after taking the limit of $E_0\rightarrow0$ and $\hat{U}_0\rightarrow0$. The difference lies in the first-law relation, where for the ultra-spinning case, it is now given by
\begin{align}
\hat{\tilde{H}}_{r_+}-\hat{H}_{r_+}&\propto\delta\hat{S}\\\nonumber&=\hat{\beta}_H(\delta\hat{M}-\hat{\Omega}_\varphi\delta\hat{J}-\hat{\Phi}\delta\hat{Q}-\hat{\Psi}\delta\hat{P}-\hat{K}\delta\lambda),
\end{align}
from eq. (\ref{firstlawultraspinning}). The last term $\hat{K}\delta\lambda$ appears from the new dynamical variable $\lambda$. However, if we treat $\lambda$ as a constant, and remain unchanged by the shock waves, we have $\hat{K}\delta\lambda=0$. The perturbation changes the black hole mass such that $\delta\hat{M}$ is also proportional to $E_0$, and for the other dynamical variables, we have $\delta\hat{J}\sim E_0\mathcal{L},\mathcal{Q}\sim\delta q/E_0,\mathcal{P}\sim\delta p/E_0$, and the first-law equation becomes
\begin{equation}
\hat{\tilde{H}}_{r_+}-\hat{H}_{r_+}\sim\hat{\beta}_H\delta \hat{M}(1-\hat{\Omega}_\varphi\mathcal{L}-\hat{\Phi}\mathcal{Q}-\hat{\Psi}\mathcal{P}).\label{HHultraspinning}
\end{equation}
In this case, from eq. (\ref{ultraspinningfunction}), $\delta\hat{M}$ is also proportional to $\lambda$, while all $\lambda$ inside the parenthesis of eq. (\ref{HHultraspinning}) cancel out. Therefore, eq. (\ref{HHultraspinning}) is also proportional to $\lambda$ and can be expressed as $\delta\hat{M}\sim\lambda E_0$. \\
\indent Since the value of $\hat{F}'(r_+)/\hat{H}'(r_+)$ also scales as the inverse of the black hole entropy $\hat{S}$ for large $\hat{S}$, i.e.
\begin{equation}
\frac{\hat{F}'(r_+)}{H'(r_+)}=\frac{\hat{\mathbb{B}}}{\hat{S}},
\end{equation}
for some $\hat{\mathbb{B}}$, the value of the shift
\begin{equation}
\alpha=\frac{\hat{\beta}_H\lambda E_0(1-\hat{\Omega}_\varphi\mathcal{L}-\hat{\Phi}\mathcal{Q}-\hat{\Psi}\mathcal{P})}{\hat{U}_0\hat{S}},
\end{equation}
does not depends on $\lambda$. This is desired because the physical scrambling time $t_*$ which is computed later on should not depend on some arbitrary parameter $\lambda$.
\section{Lyapunov Index, Scrambling Time and its Delay}\label{section4}
\subsection{Dyonic Kerr-Sen-AdS$_4$ black hole as the holographic model}
The dyonic Kerr-Sen-AdS$_4$ black hole is an asymptotically AdS rotating black hole with a complete commuting set of conserved charges given by (aside from the conserved energy/mass $M$) $J, Q,P$ given in eq. (\ref{normalfunctions}) with their corresponding chemical potentials are respectively given by $\mu,\Phi,\Psi$. This eternal black hole in AdS is similar to \cite{MaldacenaEternal2003} and, using the AdS/CFT correspondence, assumed to be dual to two copies of large-$N$ CFT in its asymptotic boundaries which are described by the thermo-field double (TFD) state
\begin{widetext}
\begin{equation}
|\psi\rangle=\frac{1}{\sqrt{Z[\beta_H,\mu,\Phi,\Psi]}}\sum_ne^{-\beta_H(H-\mu J-\Phi Q-\Psi P)}|\psi_n\rangle_L\otimes|\psi_n\rangle_R,\label{TFD}
\end{equation}
\end{widetext}
where $L$ and $R$ denote the CFT who lives in the left and right asymptotic boundaries respectively. The partition function is given by
\begin{equation}
Z[\beta_H,\mu_i]=\text{Tr}(e^{-\beta_H(H-\mu J-\Phi Q-\Psi P)}),
\end{equation}
where $\mu_i$ are the chemical potentials, and the density matrix is given by
\begin{equation}
\rho=\frac{e^{-\beta_H(H-\mu J-\Phi Q-\Psi P)}}{Z[\beta_H,\mu_i]}.
\end{equation}
The entanglement (von Neumann) entropy corresponds to the reduced density matrix from tracing out one of the asymptotic boundaries that recovers the Bekenstein-Hawking entropy.\\
\indent Chaos in the CFT can be scrutinized by the Lyapunov exponent $\lambda_L$ which appears in the decay of the out-of-time ordered correlators (OTOC) corresponding to the TFD state in eq. (\ref{TFD}),
\begin{equation}
\frac{\langle W(0)V(t)W(0)V(t)\rangle}{\langle WW\rangle\langle VV\rangle}=1-\varepsilon e^{\lambda_L t}+...,
\end{equation}
where $\varepsilon\sim 1/N$ is the perturbation parameter which is assumed to be inversely proportional to the number of degrees of freedom of the CFT $N$. Thus, the OTOC vanishes at
\begin{equation}
t_*\sim\log N,
\end{equation}
which is defined to be the scrambling time. The system admits fast scrambling for $t_*$ which depends logarithmically on $N$. The decay of OTOC can also be associated with the decay of mutual information defined as
\begin{equation}
I(A;B)=S_A+S_B-S_{A\cup B},\label{mutualinfo}
\end{equation}
where $S_A$ is the entanglement entropy associated with a reduced density matrix $\rho_A$. This can be understood from the fact that the mutual information provides an upper bound for correlations between two subsystems \cite{Wolf2008}
\begin{equation}
I(A;B)\geq\frac{(\langle\mathcal{O}_A\mathcal{O}_B\rangle-\langle\mathcal{O}_A\rangle\langle\mathcal{O}_B\rangle)^2}{2||\mathcal{O}_A||^2||\mathcal{O}_B||^2}.\label{mutualinfobound}
\end{equation}
\\
\indent The dual CFT associated with this black hole has non-trivial entanglement between the degrees of freedom in the left and right boundaries. The entanglement entropy of a sub-region in the CFT can be calculated holographically using the Ryu-Takayanagi (RT) or Hubeny-Rangamani-Takayanagi (HRT) surfaces from the AdS/CFT correspondence. The RT/HRT surface is minimal in the bulk which is homologous to the sub-region of the CFT. The entanglement entropy is then given by the area of the RT/HRT surface divided by $4G_N$. This can also be used to calculate the mutual information in eq. (\ref{mutualinfo}), and from eq. (\ref{mutualinfobound}), $I(A;B)\rightarrow0$ indicates scrambling and thus the scrambling time $t_*$ can be determined from there. In this work, we calculate $I(A;B)$ for the CFT given by the TFD state in eq. (\ref{TFD}) holographically, and extract its scrambling time $t_*$ and the Lyapunov exponent $\lambda_L$.\\
\indent The scrambling phenomena happens after we mildly perturb our TFD state at a very early time on the left asymptotic boundary. In the bulk description, the perturbation corresponds to a shock wave that propagates from the boundary to the black hole. The geometry can be understood as the Dray-'t Hooft solution. The following section shows that even an infinitesimally small perturbation can still cause scrambling, disrupting the entanglement.
\subsection{Lyapunov Exponent and Scrambling Time}
\begin{figure*}
\includegraphics[scale=0.8]{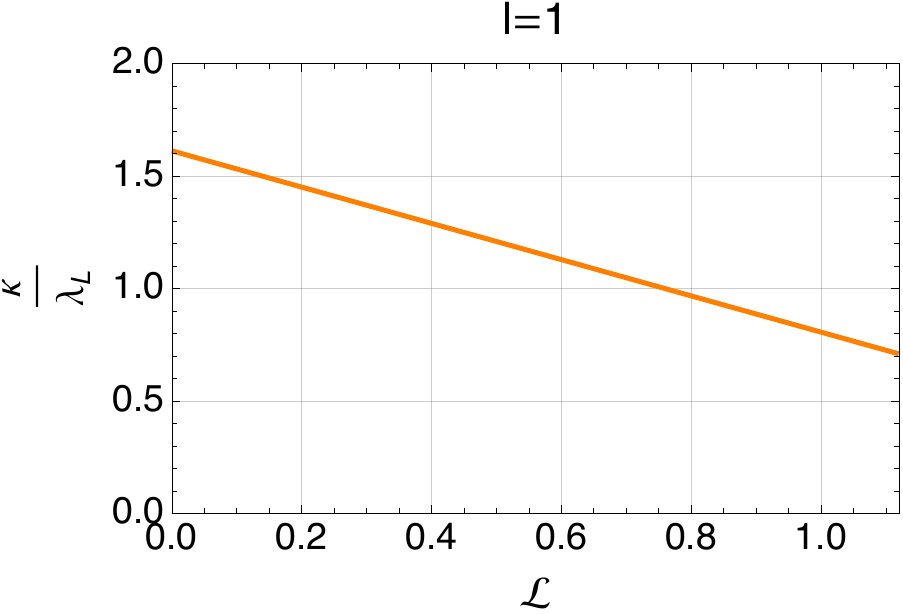}\;\;\;
\includegraphics[scale=0.6]{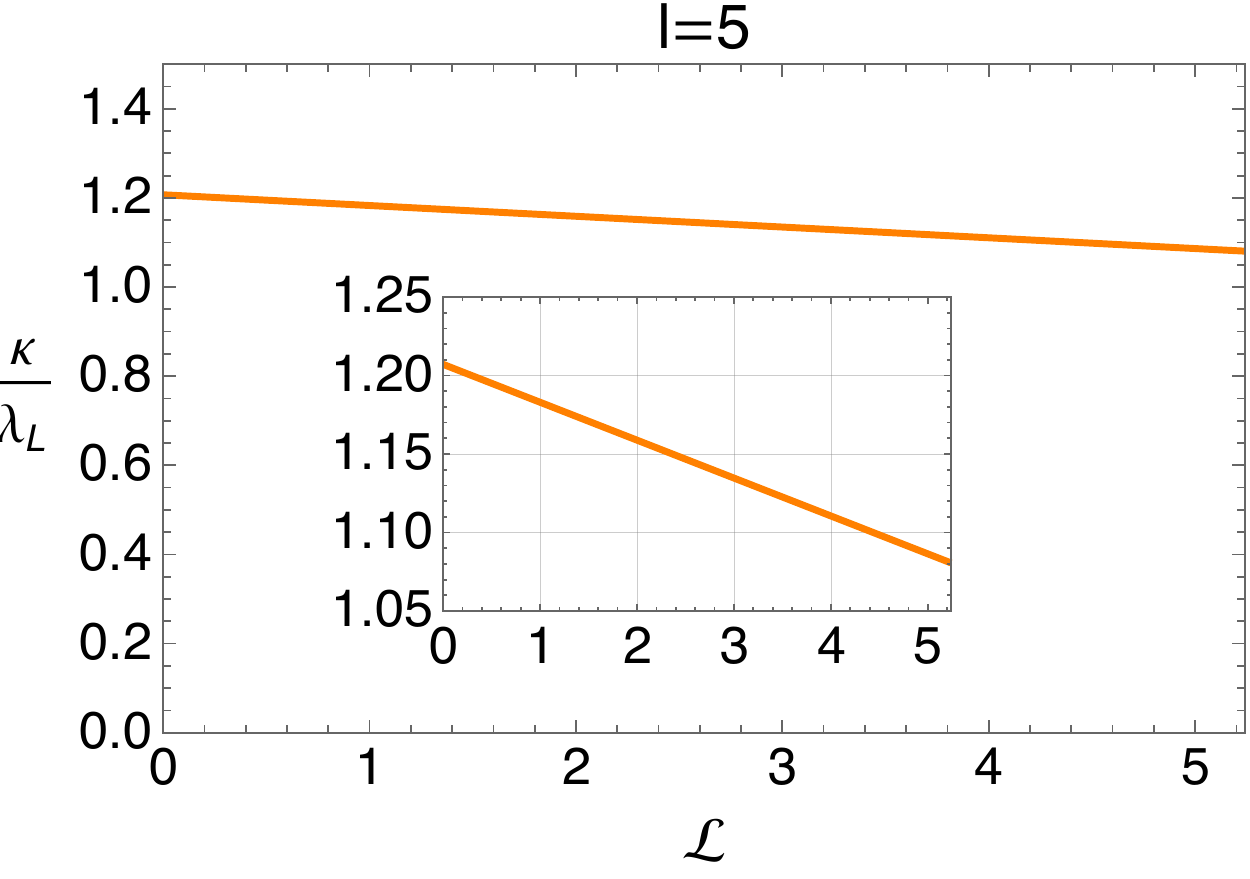}\\
\includegraphics[scale=0.6]{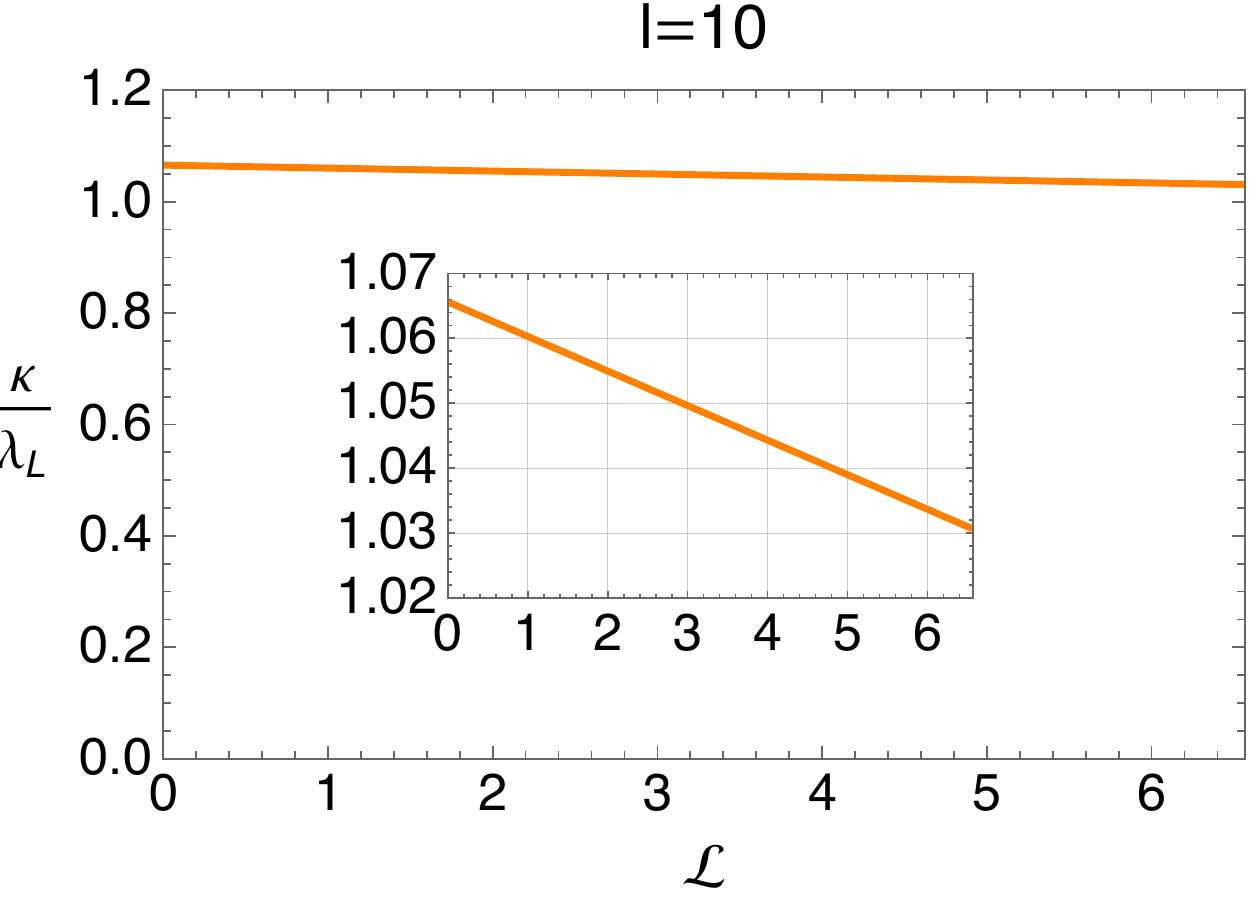}\;\;\;
\includegraphics[scale=0.6]{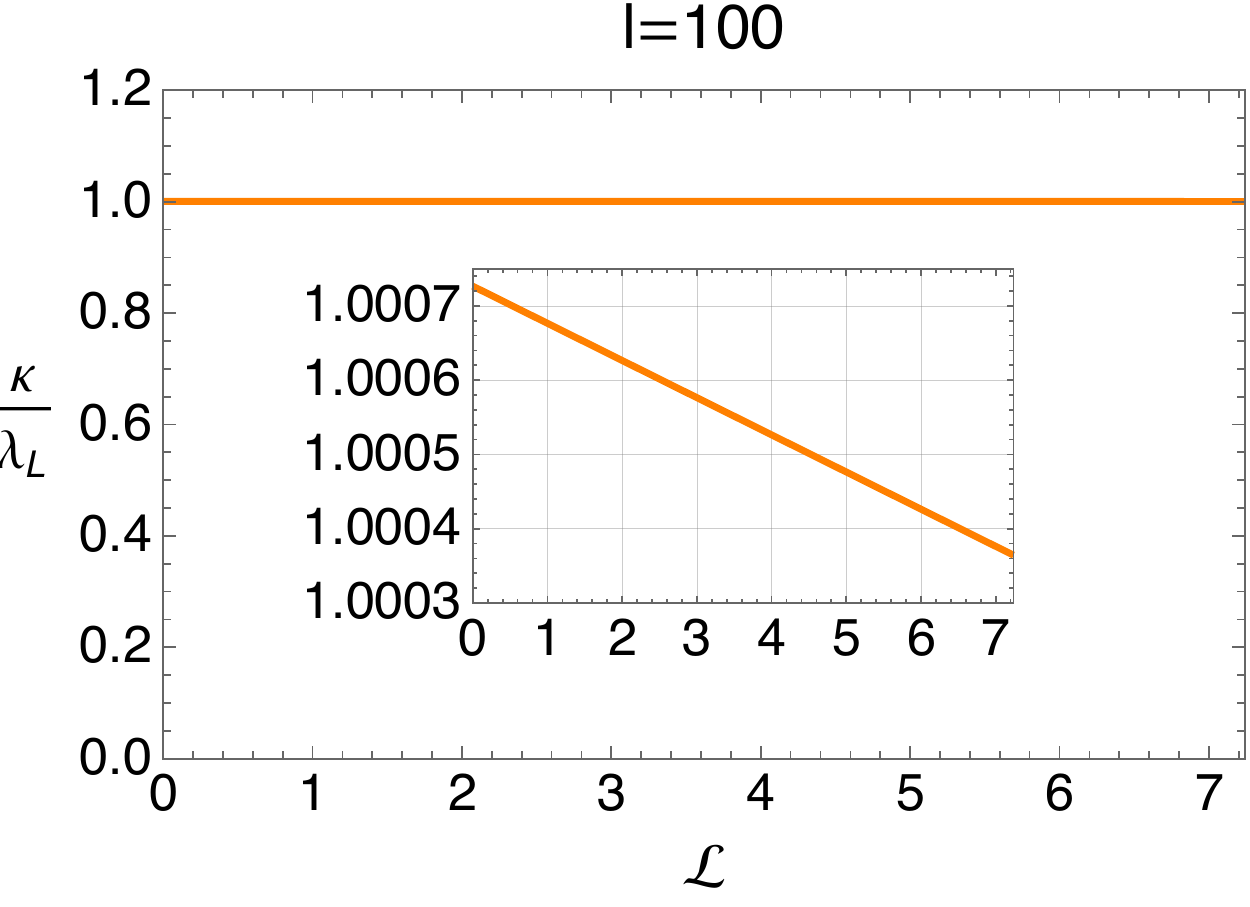}
\caption{Plot for the ratio between $\kappa$ and $\lambda_L$ with respect to $\mathcal{L}$ up to $\mathcal{L}\rightarrow1/\mu$. We use an example with $m=1,a=0.5,p=0.2,q=0.1$ and varying $l$ from $l=1$ (top-left), $l=5$ (top-right), $l=10$ (bottom-left), and $l=100$ (bottom-right). Those examples are quite far from extremality with the outer and inner horizons ratio given by $r_-/r_+=\{0.199265, 0.100242, 0.092043,0.0887473\}$ for $l=\{1, 5, 10,100\}$ respectively. For the first case with $l=1$, $\lambda_L$ and $\kappa$ intersect each other at $\mathcal{L}=0.759852$ or in another perspective $\mathcal{L}\approx0.677925/\mu$. For the other cases ($l=5,10,100$) the ratio is almost constant w.r.t $\mathcal{L}$. As $l$ gets higher, the ratio approaches 1.}\label{fig:Lyapunovnonextremal}
\end{figure*}
\begin{figure*}
\includegraphics[scale=0.8]{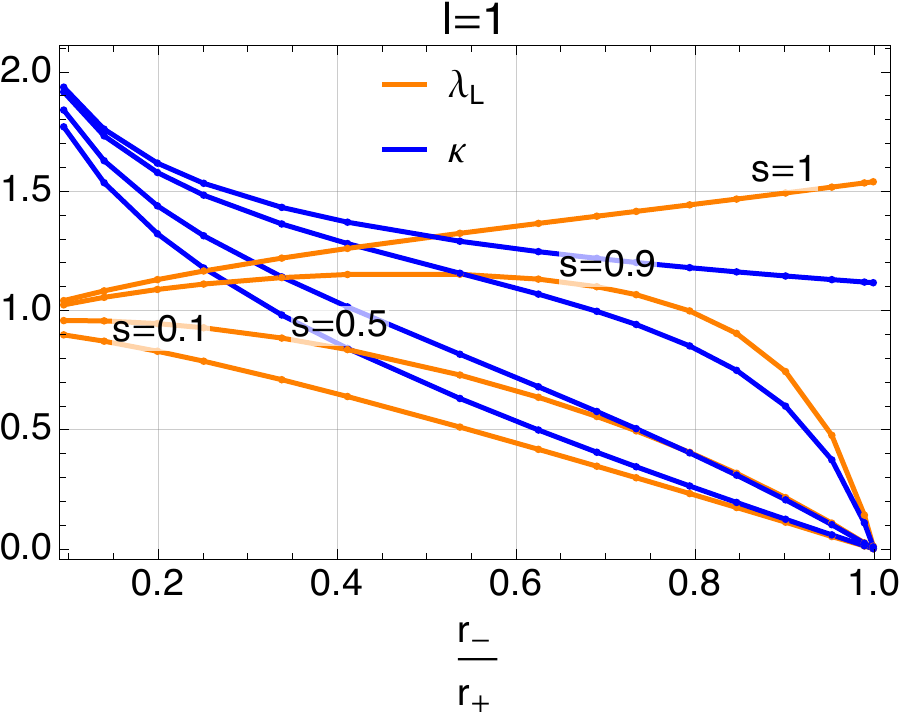}\;\;\;
\includegraphics[scale=0.8]{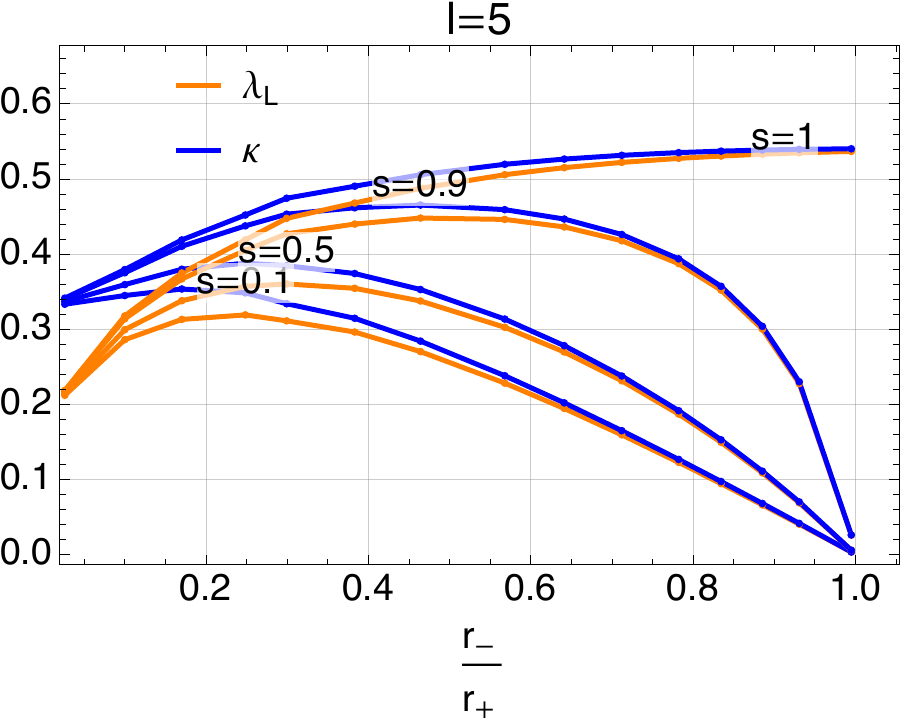}\\
\includegraphics[scale=0.8]{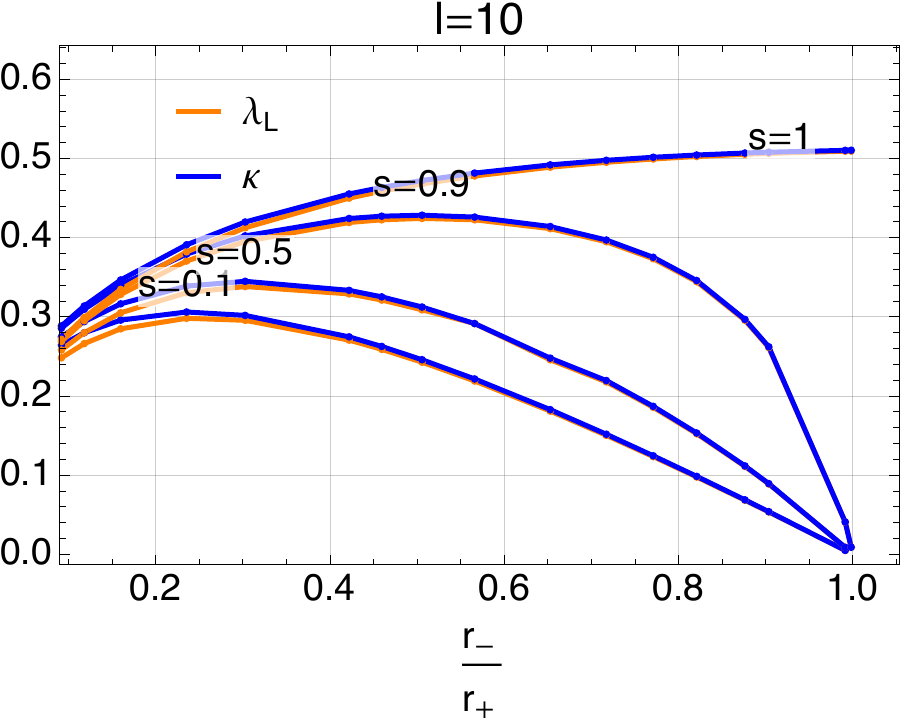}\;\;\;
\includegraphics[scale=0.8]{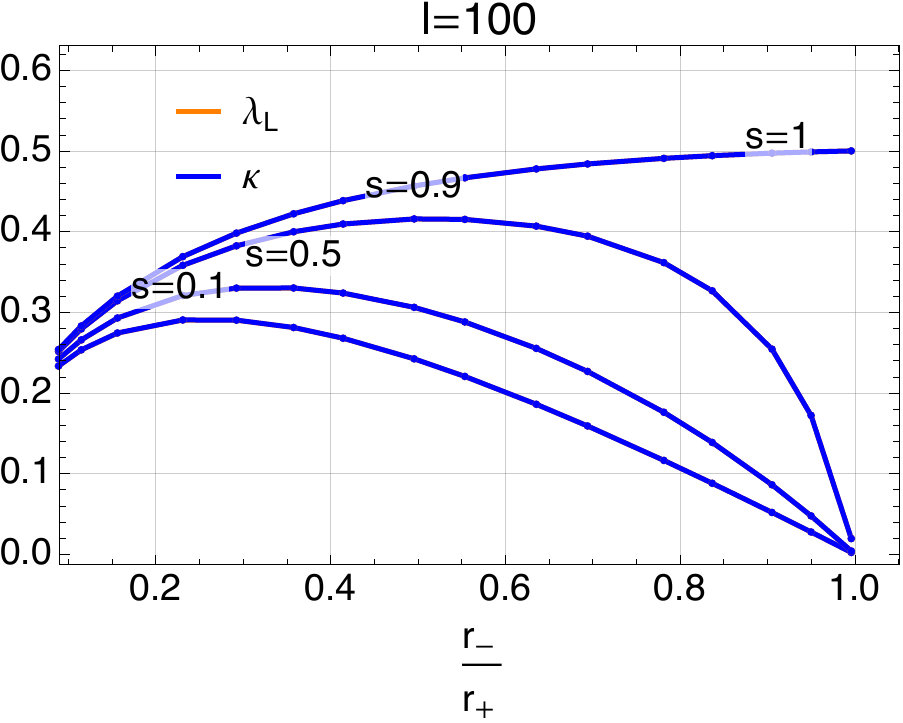}
\caption{Plot of both $\lambda_L$ and $\kappa$ with $\mathcal{L}$ is scaled as $\mathcal{L}=s\frac{r_-}{r_+}\mu^{-1}$ from $\frac{r_-}{r_+}\rightarrow0$ to $\frac{r_-}{r_+}\rightarrow1$, for $s=\{0.1,0.5,0.9,1\}$. We vary $l$ from $l=1$ (top-left), $l=5$ (top-right), $l=10$ (bottom-left), and $l=100$ (bottom-right). We gradually decrease the black hole's mass while keeping other parameters fixed: $a=0.5,p=0.2,q=0.1$. The corresponding mass are $m\in[1.5, 0.65474]$ for $l=1$, $m\in[2,0.55163]$ for $l=5$, $m\in[1,0.54729]$ for $l=10$, and $m\in[1,0.54582]$ for $l=100$.\label{fig:Lyapunovscaling}}
\end{figure*}
\begin{figure}
\includegraphics[scale=0.6]{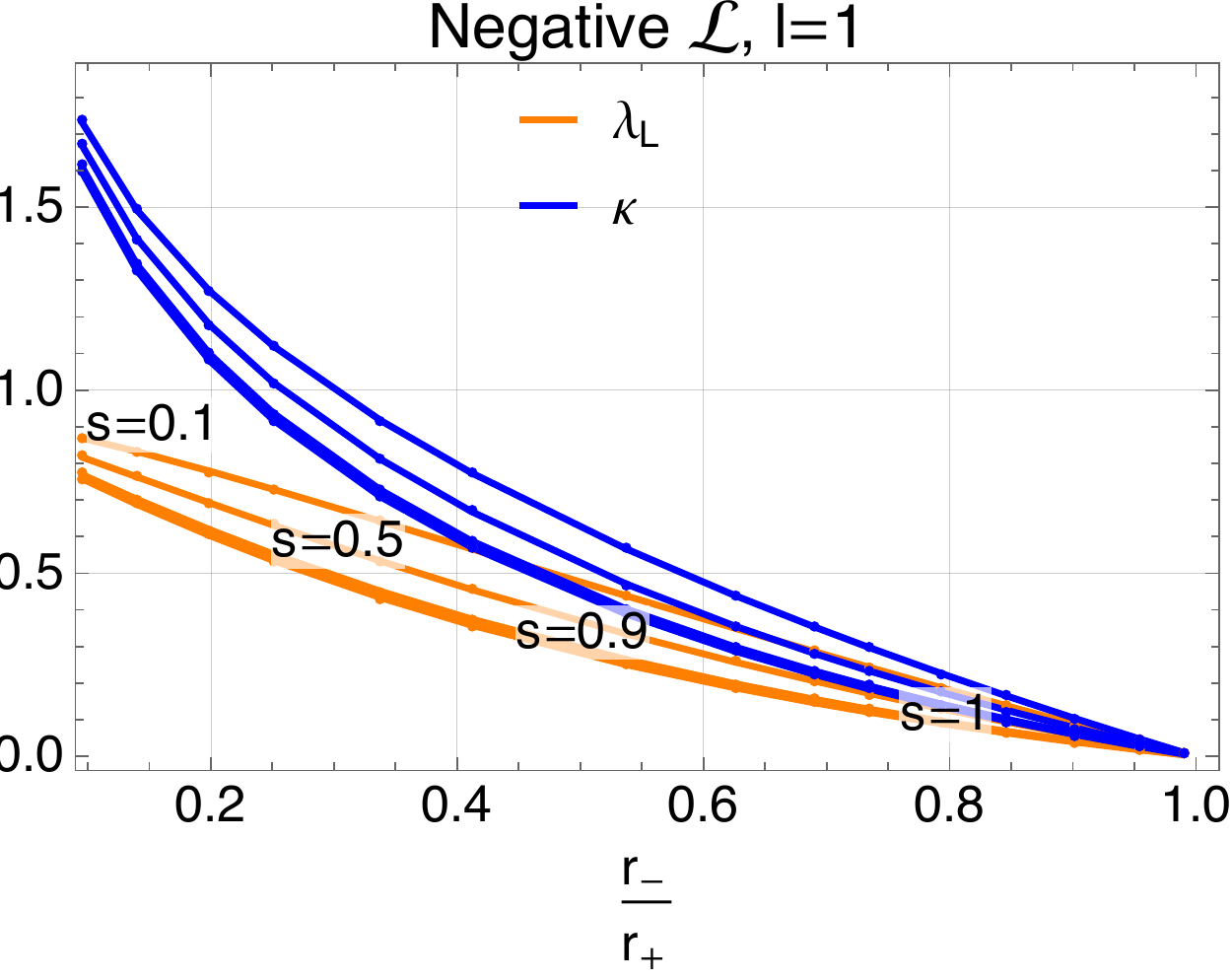}
\caption{Plot of both $\lambda_L$ and $\kappa$ for $l=1$, with remaining parameters are identical to the ones in Figure \ref{fig:Lyapunovscaling} (top-left). In this case, we reverse the sign of $\mathcal{L}$ by performing $s\rightarrow -s$.\label{fig:negativeL}}
\end{figure}

We consider two sub-regions $A$ and $B$ to be identical at their left and right asymptotic boundaries, respectively, with the equator $\theta=\pi/2$ serving as the boundary, following \cite{Malvimat2022KerrAdS}. The RT/HRT surface for both $S_A$ and $S_B$ does not depend on the Dray-'t Hooft shift $\alpha$ since they lie outside the outer horizon $r_+$. However, the RT/HRT surface corresponding to $S_{A\cup B}$ penetrates the horizon; we will call this surface $\mathcal{A}_{A\cup B}$. It has a turning point inside it, then connects to the other side of the asymptotic boundary. Thus, $\alpha$ plays an important role here. \\
\indent Due to its symmetry, the surface $\mathcal{A}_{A\cup B}$ can be obtained by extremizing the following integral
\begin{equation}
\mathcal{A}=2\pi\int d\tau\sqrt{h}\sqrt{-F+F\frac{\tilde{f}^2\dot{r}^2}{\Delta^2}}.
\end{equation}
This surface consists of three segments, I, II, and III, where I stretches from the left asymptotic boundary $(U,V)=(1,-1)$ to a point which intersects $V=0$, i.e.  $(U,V)=(U_1,0)$, II stretches from $(U,V)=(U_1,0)$ to the turning point at $r_\star$, and II stretches from the turning point to the intersection point at $U=0$ and $V=\alpha/2$. The area $\mathcal{A}_{A\cup B}$ is then equal to four times $\mathcal{A}$. Since the area functional does not depend on $\tau$, there exists a conserved quantity defined as
\begin{equation}
\frac{-F\sqrt{h}}{\sqrt{-F+F\tilde{f}^2\dot{r}^2/\Delta^2}}=\sqrt{F_\star h_\star},
\end{equation}
where $F_\star\equiv F(r_\star)$ and $h_\star\equiv h(r_\star)$ with $r_\star$ is defined as the turning point where $\dot{r}=0$ (not to be confused with the tortoise coordinate $r_*$ defined in eq. (\ref{tortoise})). \\
\indent By following similar calculations done in \cite{Malvimat2022KerrAdS} (and earlier by \cite{Leichenauer2014}), we have
\begin{equation}
\alpha=2\exp\bigg(Q_1+Q_2+Q_3\bigg),
\end{equation}
where the functions $Q_1,Q_2,Q_3$ are defined as
\begin{align}
Q_1&=-2\kappa\int_{\bar{r}}^{r_0}\frac{\tilde{f}dr}{-\Delta},\\
Q_2&=2\kappa\int_{r_\star}^\infty\frac{\tilde{f}dr}{\Delta}\bigg(1-\frac{1}{\sqrt{1+(Fh/F_\star h_\star)}}\bigg),\\
Q_3&=2\kappa\int_{r_\star}^{r_+}\frac{\tilde{f}dr}{\Delta}\bigg(1+\frac{1}{\sqrt{1+(Fh/F_\star h_\star)}}\bigg),
\end{align}
where $\bar{r}$ is defined as the location where $r_*=0$. Both $Q_1$ and $Q_2$ diverge as $r_\star\rightarrow r_+$, which correspond to $\alpha\rightarrow0$. On the other hand, $Q_3$ diverges as $r_\star\rightarrow r_c$, where $r_c$ satisfies
\begin{equation}
h(r_c)F'(r_c)+h'(r_c)F(r_c)=0.
\end{equation}
\begin{figure}
\includegraphics[scale=0.6]{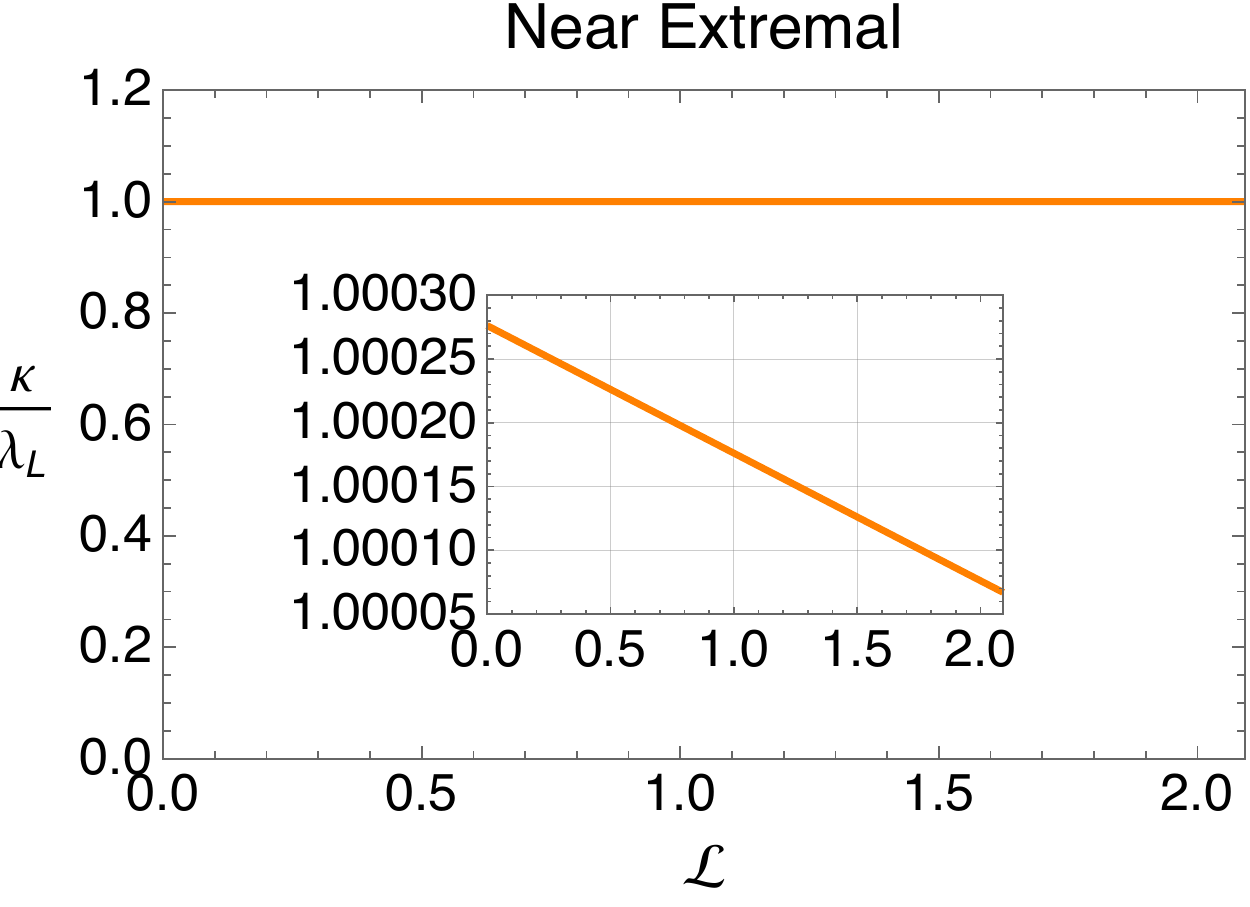}
\caption{Similar plot with Figure \ref{fig:Lyapunovnonextremal} for near-extremal case. We choose $m=1.03865,a=1,p=0.2,q=0.2$ which gives two almost degenerate horizons $r_+=1.04562$ and $r_-=1.03103$ with ratio $r_-/r_+\approx0.986$. In this case, the ratio also constant w.r.t. $\mathcal{L}$ and it is close to 1.}\label{fig:Lyapunovnearextremal}
\end{figure}
This limit corresponds to $\alpha\rightarrow\infty$, and hence it is the limit where the shift $\alpha$ becomes important. The area $\mathcal{A}_{A\cup B}$ can then be obtained to be proportional to $Q_3$, i.e.
\begin{align}
\mathcal{A}_{A\cup B}&\approx\frac{4\pi}{\kappa}\sqrt{-F(r_c)h(r_c)}Q_3\\ \nonumber&=\frac{4\pi}{\kappa}\sqrt{-F(r_c)h(r_c)}\log\alpha.
\end{align}
\indent By plugging in the value of $\alpha$ in eq. (\ref{alpha}) into $\mathcal{A}_{A\cup B}$ and use $U_0=e^{-\kappa\tau_0}$, we obtain
\begin{align}
\mathcal{A}_{A\cup B}\approx4\pi\tau_0&\sqrt{-F(r_c)h(r_c)}\\\nonumber
&+\frac{4\pi}{\kappa}\sqrt{-F(r_c)h(r_c)}\\\nonumber&\times\log\bigg(\frac{\beta_H E_0(1-\mu\mathcal{L}-\Phi\mathcal{Q}-\Psi\mathcal{P})}{S}\bigg).
\end{align}
The area $\mathcal{A}_{A\cup B}$ grows linearly in time and picks up a contribution from the perturbations $\mathcal{L},\mathcal{Q},\mathcal{P}$ in the second term. The rate of growth of $\mathcal{A}_{A\cup B}$ depends on $\sqrt{-F(r_c)h(r_c)}$, which also gives us the information about the quantum Lyapunov exponent. Following \cite{Malvimat2022KerrAdS}, the instantaneous Lyapunov exponent is given by
\begin{equation}
\lambda^{\text{(min)}}_L=\frac{4\pi\sqrt{-F(r_c)h(r_c)}}{\mathcal{A}_H},\label{LyapunovExponent}
\end{equation}
where $\mathcal{A}_H$ is the horizon's area. This Lyapunov exponent depends on the angular momentum $\mathcal{L}$, as shown in Figure \ref{fig:Lyapunovnonextremal} for some particular example with two horizons, away from extremality. This gives us enough insight that the Lyapunov exponent $\lambda_L$ is approximately equal to $\kappa$ and the ratio $\kappa/\lambda_L$ hardly depends on the angular momentum $\mathcal{L}$ of the shocks. However, for smaller $l$, $\kappa/\lambda_L$ becomes more likely to depends on $\mathcal{L}$ and for the case $l=1$, we have $\kappa/\lambda_L<1$ for $\mathcal{L}>0.759852$. In this particular case, the Lyapunov exponent seems to violate its upper bound $\kappa$ for large $\mathcal{L}$. At near-extreme conditions, the difference between $\lambda_L$ and $\kappa$ becomes even tighter, as shown in Figure \ref{fig:Lyapunovnearextremal}. This also indicates that the Lyapunov exponent of the dyonic Kerr-Sen-AdS$_4$ is also bounded by $\kappa$, and approaches its maximal value when the black hole approaches extremality.\\
\indent More detailed behaviors of the Lyapunov exponent can be seen in Figure \ref{fig:Lyapunovscaling}. Now, we scale up the angular momentum of the particle as $\mathcal{L}=s\frac{r_-}{r_+}\mu^{-1}$, following \cite{Malvimat2022KerrAdS}, so that it approaches $\mu\mathcal{L}\rightarrow s$ when the black hole becomes extremal at $\frac{r_-}{r_+}\rightarrow1$. Note that, in this paper, we work with the solution in which there are only two real and positive horizons, $r_+$ and $r_-$ with $r_+>r_-$. For $l=1$, we observe the violation of $\lambda_L\leq\kappa$ bound for large $\mathcal{L}$. Interestingly, such a violation has been observed earlier in various black holes involving charges and rotations (See, for example, \cite{Zhao2018,Gwak_2022,Yu2023Violating, Yu_2022}), for some large values of the rotation parameter. In this work, although we are already using $\kappa=\kappa_0/(1-\mu\mathcal{L})>\kappa_0$, the Lyapunov exponent can still surpass the value of $\kappa$ at large $\mathcal{L}$. This violation was not observed in the standard (uncharged) Kerr-Sen-AdS black hole \cite{Malvimat2022KerrAdS}. Such a violation might occur due to the existence of extra charges such as the dilaton and axion charges. From the plot, we also see that the Lyapunov exponent approach $\kappa$ as $l$ becomes large, i.e. the ratio $\kappa/\lambda_L$ approaches 1 in such cases. All of the values of $\lambda_L$ and $\kappa$ approach zero as we reach extremality $\frac{r_-}{r_+}$ since $T_H\rightarrow0$ at this limit. There is an exception for $s=1$, since $\mathcal{L}\rightarrow\mu^{-1}$ at extremality and $\kappa$ becomes $\kappa_{\text{ext.}}$. This feature is interesting in studying the behavior of chaos in extremal black holes \cite{Banerjee2020}.\\
\indent The violation of the standard MSS chaos bound in the EMDA theory has been investigated earlier in \cite{Yu_2022} using particle's geodesic. They find that the violation is more likely when the particle rotates in the opposite direction from the black hole. Using our analysis, we also see how the Lyapunov exponent behaves as we change the sign of $\mathcal{L}$ to its negative counterpart. The result can be seen in Figure \ref{fig:negativeL}. However, in this case, the violation does not occur. For the negative value of $\mathcal{L}$, all $\lambda_L$ and $\kappa$ approach zero as the black hole approaches extremality. This is because the numerator $(1-\mu\mathcal{L})$ will never reach zero, while the temperature of the black hole approaches zero at extremality. We suggest that the chaotic behavior with a negative value of $\mathcal{L}$ is also crucial in understanding the chaotic behavior of an extremal black hole. The behavior of $\lambda_L$ and $\kappa$ at extremality, whether they are zero or not, depends on the sign of $\mathcal{L}$. This needs to be investigated further.\\
\indent The violation of the bound in our case can be understood from the behavior of the black hole's entropy for large and small $l$. For small $l$ with $l\approx a$, the temperature becomes
\begin{equation}
2\pi T_H\approx\frac{2 r_+}{l^2}-\frac{M}{r_+^2-d^2-k^2+l^2}.
\end{equation}
Here, the role of the dilaton and axion charges $d,k$ is to lower the temperature, and hence, also lower the value of $\kappa$. On the other hand, for large $l$ with $l\gg a$, we have
\begin{equation}
2\pi T_H=\frac{r_+-M}{r_+^2-d^2-k^2+a^2}.
\end{equation}
In contrast with the previous case, here, the dilaton and axion charges enlarge the black hole's temperature. For the black hole with non vanishing $d,k$ we expect lower value of $\kappa$ for smaller $l$, and hence the violation of the chaos bound may occur.\\
\indent From the formula of the Lyapunov exponent given by eq. (\ref{LyapunovExponent}) alone, it is hard to see whether the violation of the $\kappa$ bound for a small value of $l$ is caused by the dilaton or axion charges. However, this might be physically interesting since such a violation does not occur in other analogous black holes such as the standard Kerr-AdS$_4$ calculations \cite{Malvimat2022KerrAdS}. Conducting further investigations on different types of black holes could be crucial in gaining insights that allow us to pinpoint the source of the violation.
\subsubsection*{Ultra-spinning Case}
\begin{figure}
\includegraphics[scale=0.6]{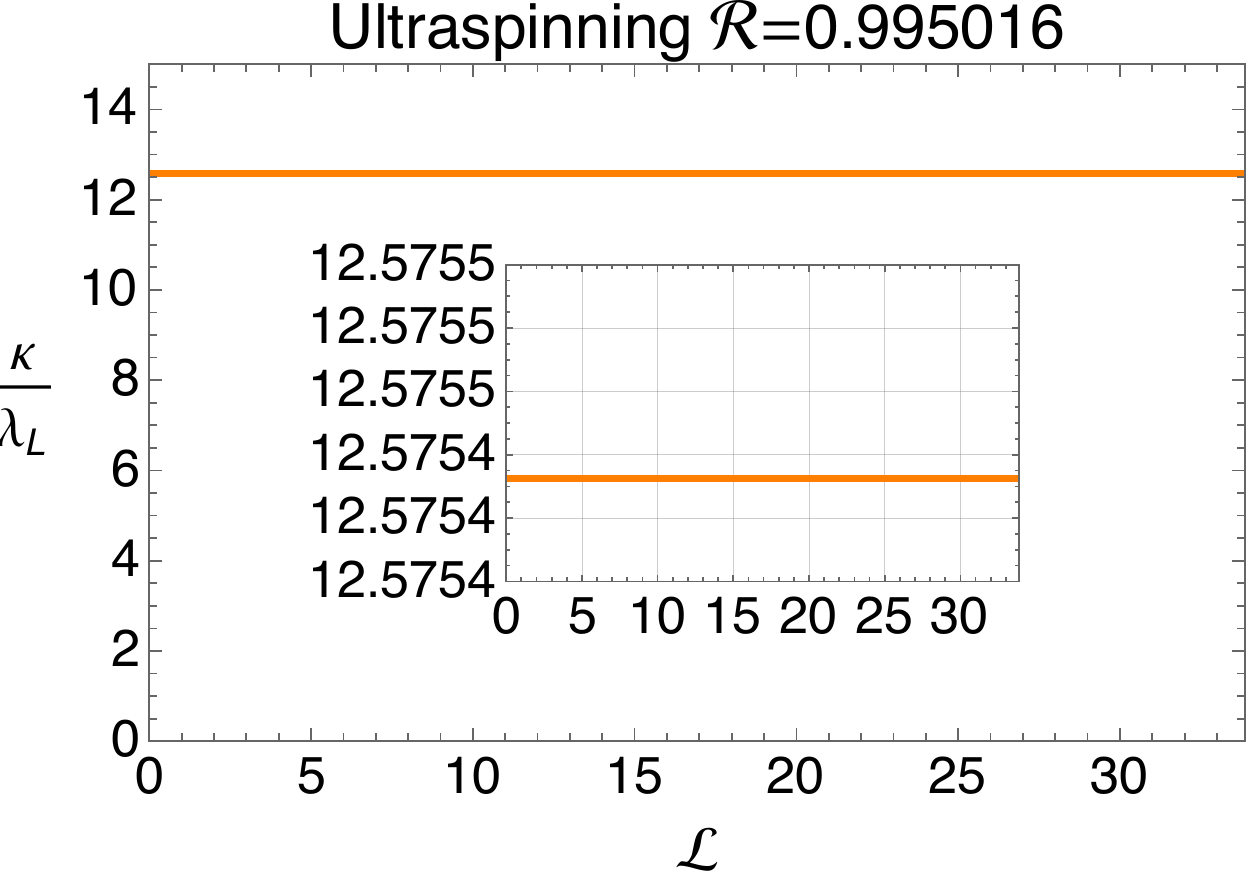}\;\;\;
\includegraphics[scale=0.6]{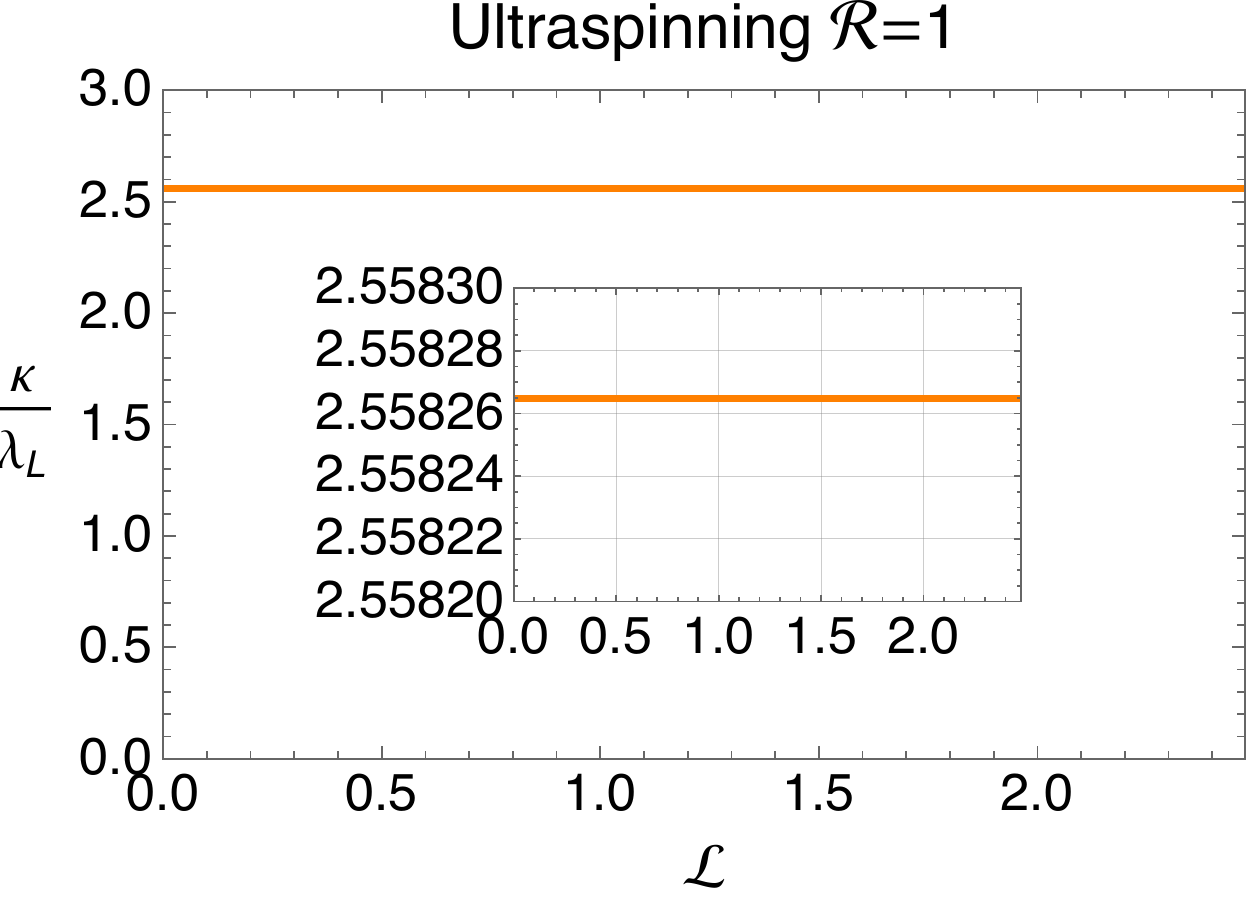}
\caption{Plot similar to Figure \ref{fig:Lyapunovnonextremal} for the ultra-spinning cases. We use $m=100,p=0.2,q=0.1,l=1$ for the first case (top) which violates the RII, $\mathcal{R}=0.995016$. For the second case (bottom), we use $m=5,p=2,q=1,l=0.5$ which does not violate the RII, $\mathcal{R}=1$. Both cases are non-extremal with $r_-/r_+=\{0.000915922,0.478328\}$ for the first and the second cases respectively.\label{fig:plotultraspin}} 
\end{figure}
For the ultra-spinning case, we should be careful when integrating the transverse volume since now $\phi$ takes arbitrary periodicity, which is denoted by $\lambda$. In our derivation around eq. (\ref{etaultraspinning}), we define $\hat{\eta}$ such that we recover the horizon area when we integrate $\hat{z}$ from 0 to $\lambda$. Therefore, the area functional in the ultra-spinning case should be written as
\begin{equation}
\mathcal{A}=\lambda\int d\tau\sqrt{\hat{h}}\sqrt{-\hat{F}+\hat{F}\frac{\hat{\tilde{f}}^2\dot{r}^2}{\hat{\Delta ^2}}}.
\end{equation}
Following similar derivations as before, we obtain the area $\mathcal{A}_{A\cup B}$ in the limit of $r_\star\rightarrow r_c$ which corresponds to large $Q_3$. The value of $r_c$ can be obtained by solving
\begin{equation}
\hat{h}(r_c)\hat{F}'(r_c)+\hat{h}'(r_c)\hat{F}(r_c)=0.
\end{equation}
Thus, the area $\mathcal{A}_{A\cup B}$ is again proportional to $Q_3$:
\begin{align}
\mathcal{A}_{A\cup B}&\approx\frac{2\lambda}{\kappa}\sqrt{-\hat{F}(r_c)\hat{h}(r_c)}Q_3\\\nonumber&=\frac{2\lambda}{\kappa}\sqrt{-\hat{F}(r_c)\hat{h}(r_c)}\log\hat{\alpha}.
\end{align}
\indent Although the area $\mathcal{A}_{A\cup B}$ depends on $\lambda$, interestingly, the minimal instantaneous Lyapunov index $\lambda_L^{\text{(min)}}$ does not depend on $\lambda$. This is so because the black hole horizon area $\mathcal{A}_H$ also depends linearly on $\lambda$ and the two cancel each other. For the ultra-spinning case, we can write the Lyapunov index as
\begin{equation}
\lambda_L^{(\text{min})}=\frac{\sqrt{-\hat{F}(r_c)\hat{h}(r_c)}}{(r_+^2-d^2-k^2+l^2)}.
\end{equation}
We can then study the behavior of $\lambda_L^{(\text{min})}$ using graphs for the ultra-spinning black hole and compare it with the value of $\hat{\kappa}$. \\
\indent It is known that for ultra-spinning black holes, there exist super-entropic cases which violate the Reverse Isoperimetric Inequality (RII) \cite{Hennigar2015}, which is given by
\begin{equation}
\mathcal{R}=\bigg(\frac{r_+^2}{r_+^2-d^2-k^2+l^2}\bigg)^{1/6}\geq1,
\end{equation}
for the dyonic Kerr-Sen-AdS$_4$ black hole \cite{Wu2020,Sakti2022KerrSen,Sakti2023}. An ultra-spinning dyonic Kerr-Sen-AdS$_4$ black hole always violates the RII for $0\leq q^2+p^2< 2ml$ or $0\leq d^2+k^2<l^2$. However, for other values of $q,p$ (or equivalently $d,k$), we can have cases in which the RII is not violated. In this work, we see the behavior of $\lambda_L^{(\text{min})}$ and $\hat{\kappa}$ for two cases: obeys RII and violates RII. The result is shown in Figure \ref{fig:plotultraspin}. It is interesting that for both cases, $\lambda_L^{(\text{min})}$ is bounded by $\kappa$ and the ratio $\kappa/\lambda_L^{(\text{min})}$ does not depends on $\mathcal{L}$ up to the fourth decimal places. For the case which violates the RII, the gap between $\kappa$ and $\lambda_L^{(\text{min})}$ is wider than the one which does not violate the RII. It is interesting to learn more about the relationship between the Lyapunov index and the violation of the RII for ultra-spinning black holes.\\
\begin{figure*}
\includegraphics[scale=0.78]{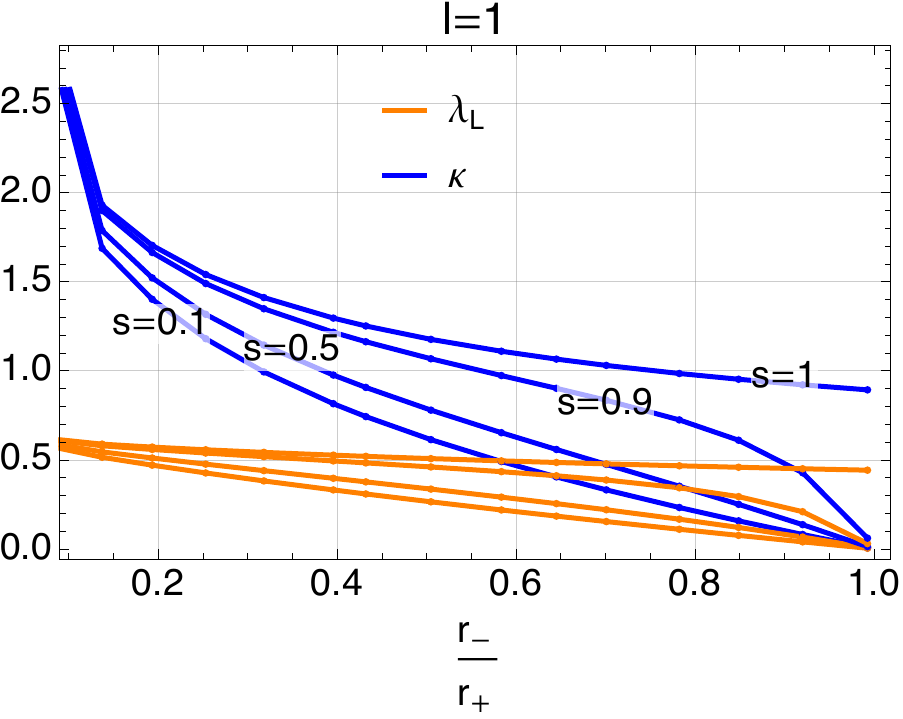}\;\;\;
\includegraphics[scale=0.8]{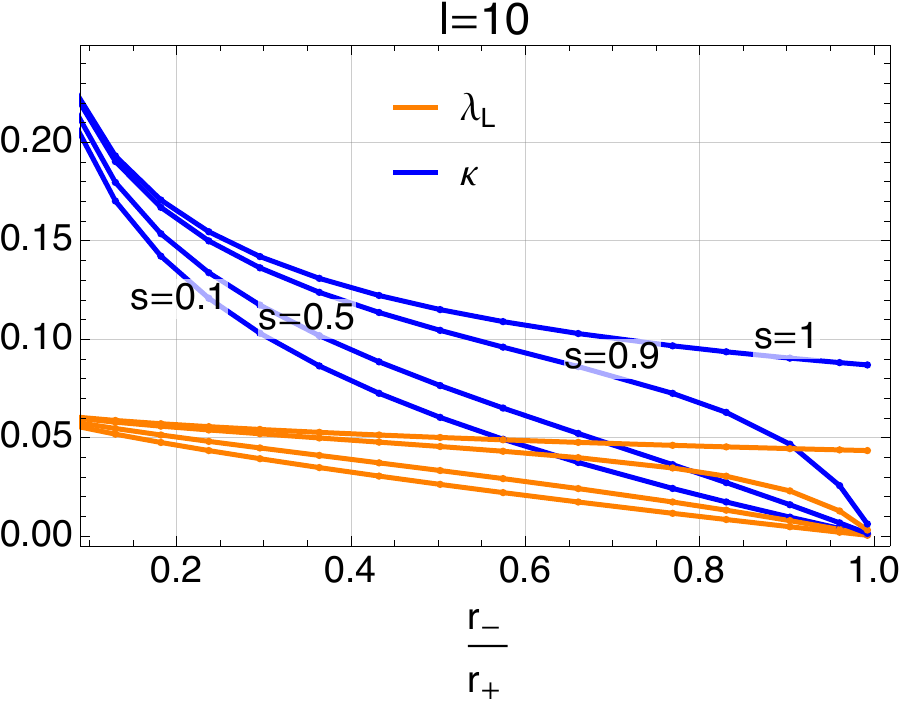}
\caption{Plot similar to Figure \ref{fig:Lyapunovscaling} for ultra-spinning black hole. The fixed parameters are $a=0.5, p=0.2,q=0.1$ while decreasing the mass to approach extremality.}\label{fig:lambdaultraspinning}
\end{figure*}
\indent After we obtain the area connecting two boundaries $\mathcal{A}_{A\cup B}$, now we are ready to calculate the Mutual Information:
\begin{align}
I(A;B)\approx &S_A+S_B-\frac{\pi\tau_0}{G_N}\sqrt{-F(r_c)h(r_c)}\\\nonumber&-\frac{\pi}{\kappa G_N}\sqrt{-F(r_c)h(r_c)}\\\nonumber&\times\log\bigg(\frac{\beta_H E_0(1-\mu\mathcal{L}-\Phi\mathcal{Q}-\Psi\mathcal{P})}{S}\bigg).
\end{align}
The scrambling time $\tau_*$ obtained from $I(A;B)\rightarrow0$ is then given by
\begin{align}\label{scramblingtimenormal}
\kappa\tau_*^{(\mathcal{L},\mathcal{Q},\mathcal{P})}\approx&\log S+\frac{\kappa(\mathcal{A}_A+\mathcal{A}_B)}{4\pi\sqrt{-F(r_c)h(r_c)}}\\\nonumber&+\log\frac{1}{1-\mu\mathcal{L}-\Phi\mathcal{Q}-\Psi\mathcal{P}},
\end{align}
while the ultra-spinning version, following a similar derivation, is given by
\begin{align}
\label{scramblingtimeultra}
\hat{\kappa}\hat{\tau}_*^{(\mathcal{L},\mathcal{Q},\mathcal{P})}\approx&\log\frac{\hat{S}}{\lambda}+\frac{\hat{\kappa}(\mathcal{A}_A+\mathcal{A}_B)}{4\pi\sqrt{-\hat{F}(r_c)\hat{h}(r_c)}}\\\nonumber&+\log\frac{1}{1-\hat{\Omega}_\varphi\mathcal{L}-\hat{\Phi}\mathcal{Q}-\hat{\Psi}\mathcal{P}}. 
\end{align}
For a black hole with large entropy, the first term is substantially larger than the rest and thus the scrambling time is approximately given by
\begin{equation}
\tau_*^{(\mathcal{L},\mathcal{Q},\mathcal{P})}\sim\frac{1}{\kappa}\log S,\;\;\;\hat{\tau}_*^{(\hat{\mathcal{L}},\hat{\mathcal{Q}},\hat{\mathcal{P}})}\sim\frac{1}{\hat{\kappa}}\log\frac{\hat{S}}{\lambda}.\label{scramblingtime}
\end{equation}
This indicates that the dyonic Kerr-Sen-AdS$_4$ and its ultra-spinning counterpart also follow the fast scrambling conjecture \cite{Sekino2008}. For the ultra-spinning case, $\hat{\tau}_*$ actually does not depends on $\lambda$ because the entropy $\hat{S}$ is proportional to $\lambda$. For an ultra-spinning black hole with large entropy $\hat{S}\rightarrow\infty$ while keeping the parameter $\lambda$ fixed, the term proportional to $\log\lambda^{-1}$ can be ignored, and hence the parameter $\lambda$ does not contribute to the scrambling time.\\
\indent However, despite being much smaller than the leading $\log S$ term, the last term of eq. (\ref{scramblingtimenormal}) and eq. (\ref{scramblingtimeultra}) provide insight into the delay of the scrambling time compared to the standard scrambling time when $\mathcal{Q},\mathcal{P}=0$. Such a delay was first investigated by \cite{horowitz_bouncing_2022} for a non-rotating charged black hole. This term does depend on the thermodynamic potentials (or chemical potentials) of the black hole: $\mu,\Phi,\Psi$, and the external perturbations $\mathcal{Q},\mathcal{P}$. If the trajectory of the shock waves bounces near the horizon, this term can be large. This will be further explored in the following subsection.\\
\indent The second term of eq. (\ref{scramblingtime}) does not scale with entropy and hence it is non-extensive since both $\mathcal{A}_A+\mathcal{A}_B$ and $\sqrt{-F(r_c)h(r_c)}/\kappa$ have dimensions of the area. Furthermore, it also not depends on the perturbations $\mathcal{Q},\mathcal{P}$ and also the plot in Figures \ref{fig:Lyapunovnonextremal} and \ref{fig:Lyapunovnearextremal} imply that it is also approximately constant in $\mathcal{L}$, at least for large value of $l$. For the ultra-spinning case, those arguments also hold (see Figure \ref{fig:plotultraspin}). Therefore, the second term, along with the first term, will not contribute to the scrambling delays.
\subsection{Scrambling time delays of the dyonic Kerr-Sen-AdS$_4$}
In this section, we investigate the phenomena of scrambling delays in the dyonic Kerr-Sen-AdS$_4$ and its ultra-spinning counterpart. Scrambling time delays come from the difference between the scrambling time from (electrically and magnetically) charged and rotating shock waves $\tau_*^{(\mathcal{L},\mathcal{Q},\mathcal{P})}$ and the standard scrambling time from rotating neutral shock waves $\tau_*^{(\mathcal{L},0,0)}$, or in other words,
\begin{align}
\Delta\tau_*&\equiv\tau_*^{(\mathcal{L},\mathcal{Q},\mathcal{P})}-\tau_*^{(\mathcal{L},0,0)}\\\nonumber&=\frac{1}{\kappa}\log\frac{1-\mu\mathcal{L}}{1-\mu\mathcal{L}-\Phi\mathcal{Q}-\Psi\mathcal{P}},\label{scramblingtimedifference}\\
\Delta\hat{\tau}_*&\equiv\hat{\tau}_*^{(\mathcal{L},\mathcal{Q},\mathcal{P})}-\hat{\tau}_*^{(\mathcal{L},0,0)}\\\nonumber&=\frac{1}{\hat{\kappa}}\log\frac{1-\hat{\Omega}_\varphi\mathcal{L}}{1-\hat{\Omega}_\varphi\mathcal{L}-\hat{\Phi}\mathcal{Q}-\hat{\Psi}\mathcal{P}},
\end{align}
for the dyonic Kerr-Sen-AdS$_4$ and its ultra-spinning counterpart respectively. From the first law of black hole thermodynamics, a positive $\delta S$ implies that the value within the logarithm is greater than one. Therefore, $\Delta\tau$ is positive, which means that it prolongs the scrambling time. We do not compare $\tau_*^{(\mathcal{L},\mathcal{Q},\mathcal{P})}$ with $\tau_*^{(0,0,0)}$ since $\kappa$ depends on $\mathcal{L}$ and therefore $\Delta\tau_*$ (and $\Delta\hat{\tau}_*$) becomes zero already as $\mathcal{Q},\mathcal{P}\rightarrow0$ even though $\mathcal{L}$ is non zero. \\
\indent In \cite{horowitz_bouncing_2022}, the time difference corresponds to the delay of the start of the scrambling process, or in other words, the scrambling delay.  They found that the time difference (similar to $\Delta\tau$) for a Reissner-Nordstr\"om black hole is equal to the difference between the bounce time $t_b$ and the time when the shell is sent from the left boundary $t_{wL}$, for time difference much larger than the thermal time $\beta$. Using the quantum circuit description, the time difference $t_d=-t_{wL}-t_b$ is then interpreted as the scrambling delay in \cite{horowitz_bouncing_2022}.\\
\indent In this work, we show that $\Delta\tau_*$ and $\Delta\hat{\tau}_*$ also corresponds to the scrambling delays for the dyonic Kerr-Sen-AdS$_4$ black hole and its ultra-spinning counterpart. We begin by calculating the stress tensor for the rotating charged shockwaves using the null junction formalism \cite{Barrabes1991, Poisson2002}, which is given by (Appendix A for derivation):
\begin{equation}
S^{\mu\nu}\propto[\sigma(r)] k^\mu k^\nu,\label{stresstensor}
\end{equation}
up to some proportionality constant, where
\begin{equation}
[\sigma(r)]\equiv\sigma_R(r)-\sigma_L(r),
\end{equation}
with the $L$ and $R$ indexes correspond to the function in the perturbed left region with parameters $M+\delta M,q+\delta q, p+\delta p, J+\delta J$ and unperturbed right region with original parameters $M,q,p,J$. For the dyonic Kerr-Sen-AdS$_4$ black hole, $\sigma_i(r)$ is given by
\begin{equation}
\sigma_i(r)=\frac{\Delta_i(r)h'_i(r)}{\tilde{f}_i(r)},\label{sigmadefinition}
\end{equation}
with $i=L,R$. If $[\sigma(r)]<0$, the null energy condition is violated. Therefore, the location of the bounce, $r_b$ can be obtained from $[\sigma(r_b)]=0$.\\
\indent Although it is difficult to exactly solve $[\sigma(r_b)]=0$, we show that we do not need the explicit form of $r_b$ to find $\tau_b$ (the "bounce time"). The increase of the black hole parameters $\delta M,\delta Q,\delta P,\delta J$ is directly related to the increase of the black hole entropy $\delta S$ by the thermodynamic first-law relation. To make sure that the perturbations increase the black hole entropy according to the second law, $\delta S>0$, thus the perturbations also increase the horizon's radius as $r_+\delta r_+ \sim\delta S $. For small perturbation parameters $\delta M,\delta q,\delta p,\delta J$, to the first order approximation, $[\sigma(r_b)]=0$ can be written as
\begin{equation}
\sigma(S+\delta S)-\sigma(S)=0\;\;\;\Longrightarrow\;\;\;\frac{\partial\sigma}{\partial S}\delta S=0.\label{deltarb}
\end{equation}
\indent We are interested in the regime where $\Delta\tau_*$ is particularly large, much larger than the thermal time $\beta\sim\frac{1}{\kappa}$. This can be achieved when the bounce happens very close to the horizon, or $r_b\approx r_+$. Therefore, eq. (\ref{deltarb}) can be expanded near $r_+$ such as
\begin{equation}
\bigg(\frac{\partial\sigma}{\partial S}\delta S\bigg)\bigg|_{r_+}+(r_b-r_+)\frac{d}{d r_b}\bigg(\frac{\partial\sigma}{\partial S}\delta S\bigg)\bigg|_{r_+}=0.\label{deltaexpandrb}
\end{equation}
Using eq. (\ref{sigmadefinition}) and the horizon's temperature formula, the first term of eq. (\ref{deltaexpandrb}) can be written as
\begin{align}
\bigg(\frac{\partial\sigma}{\partial S}\delta S\bigg)\bigg|_{r_+}&=\frac{\partial\sigma}{\partial r_b}\frac{\partial r_b}{\partial S}\delta S\bigg|_{r_+}\\\nonumber&=\frac{2\Xi}{r_+}\frac{h'(r_+)}{1-\mu\mathcal{L}}T_H\delta S,
\end{align}
which has a dimension of $r_+$. Therefore, $\frac{d}{d r_b}\bigg(\frac{\partial\sigma}{\partial S}\delta S\bigg)\bigg|_{r_+}$ in the second term of eq. (\ref{deltaexpandrb}) is dimensionless. We may extract the location of the bounce $r_b$ to be (approximately)
\begin{equation}
r_b-r_+=\tilde{\mathbb{R}}T_H\delta S=\mathbb{R}(1-\mu\mathcal{L}-\Phi \mathcal{Q}-\Psi\mathcal{P}),\label{rbrplus}
\end{equation}
where $\mathbb{R}$ has a dimension of $r_+$ and in the order of $\mathcal{O}(\delta S)$. We see that $r_b\rightarrow r_+$ corresponds to setting the condition such that $(1-\mu\mathcal{L}-\Phi\mathcal{Q}-\Psi\mathcal{P})\rightarrow0$. From eq. (\ref{scramblingtimedifference}), setting the bounce near the horizon corresponds to enlarging the scrambling time difference $\Delta \tau_*$.\\
\indent Using the Kruskal coordinate relation evaluated at $r_b$, we have
\begin{equation}
U_bV_b=e^{2\kappa r_*(r_b)},\label{ubvb}
\end{equation}
However, since we set the bounce to be close to the horizon, we have
\begin{equation}
r_*(r_b)\approx\frac{1}{2\kappa}\ln\bigg|\frac{r_b-r_+}{\mathbb{R}'}\bigg|+C,\label{rstarrb}
\end{equation}
with $\mathbb{R}'$ is some constant of dimension $r_+$ and $C$ is a dimensionless constant that depends on the black hole geometry.  Also, for $r_b\approx r_+$, we can approximate $U_b\approx U_0$, where $U_0$ is the location where the original rotating shock wave is sent. Using equations (\ref{rbrplus}), (\ref{ubvb}), (\ref{rstarrb}), as well as the relation between $U_0$ and $\alpha$, we obtain
\begin{equation}
V_b\approx\frac{\alpha S}{\beta_H E_0}\frac{\mathbb{R}}{\mathbb{R}'}e^C,\label{vb}
\end{equation}
where the factor $\frac{\mathbb{R}}{\mathbb{R}'}e^C$ is an $\mathcal{O}(1)$ dimensionless parameter which depends on the black hole's geometry and does not scale with $S$ or $\delta S$. We see that $V_b\gg1$ since $\alpha\sim\mathcal{O}(1)$ and $\beta_H E_0/S\rightarrow0$. \\
\indent Again, by plugging in the value of $\alpha$ in eq. (\ref{alpha}) into eq. (\ref{vb}), we obtain the time delay $\tau_d=\tau_0-\tau_b$ for dyonic Kerr-Sen-AdS$_4$ black hole, which is given by
\begin{align}
\tau_d=&\frac{1}{\kappa}\log\bigg(\frac{1}{1-\mu\mathcal{L}-\Phi\mathcal{Q}-\Psi\mathcal{P}}\bigg)\\\nonumber&+\frac{1}{\kappa}\log\bigg(\frac{\mathbb{R}'}{\mathbb{R}}e^{-C}\bigg),\\\nonumber
=&\Delta\tau_*+\frac{1}{\kappa}\log\bigg(\frac{1}{1-\mu\mathcal{L}}\bigg)+\frac{1}{\kappa}\log\bigg(\frac{\mathbb{R}'}{\mathbb{R}}e^{-C}\bigg).
\end{align}
Since $r_b\approx r_+$, the first term dominates all the remaining terms and we see that $\tau_d$ also approximately equals to $\Delta\tau_*$ for the dyonic Kerr-Sen-AdS$_4$ black hole. The remaining terms are in the order of thermal time $\beta\sim\frac{1}{\kappa}$ and do not scale like $\Delta\tau_*$ if we take $r_b\approx r_+$. If we set $\mathcal{L}\rightarrow0$, i.e. for the non-rotating shock waves, the result reduces to the scrambling delay found in \cite{horowitz_bouncing_2022}. This indicates that the role of the charges (electric and magnetic) in the dyonic Kerr-Sen-AdS$_4$ black hole is more to delay the start of the scrambling, rather than to prolong the scrambling process, at least for $r_b\approx r_+$. For the ultra-spinning black hole, the scrambling delay is also similar but with all of the parameters hatted:
\begin{equation}
\hat{\tau}_d=\Delta\hat{\tau}_*+\frac{1}{\hat{\kappa}}\log\bigg(\frac{1}{1-\hat{\Omega}_\varphi\mathcal{L}}\bigg)+\frac{1}{\hat{\kappa}}\log\bigg(\frac{\hat{\mathbb{R}}'}{\hat{\mathbb{R}}}e^{-\hat{C}}\bigg).
\end{equation}
\indent With vanishing $\mathcal{Q},\mathcal{P}$, only the terms with $\sim\mathcal{O}(\beta)$ remain. Away from extremality, these terms are small. However, once we approach the extremal limit, the second and the third terms of $\tau_d$ may diverge, and therefore $\Delta\tau_*$ can only be interpreted as the scrambling delay away from extremality. Furthermore, it is interesting to ask whether the bounce still happens for $\mathcal{Q},\mathcal{P}\rightarrow0$ but we keep $\mathcal{L}$. From the previous analysis, $\tau_d$ becomes very small for small $\beta$ which means that we can send the perturbation from the right asymptotic at the time close to $\tau_0$ and still meet the left perturbations inside the horizon. This indicates that the scrambling process starts immediately and the delay almost did not happen. Therefore, the charges $\mathcal{Q},\mathcal{P}$ are the ones that play the important role in delaying the scrambling process.
\section{Conclusions}\label{section5}
We study chaos in the dyonic Kerr-Sen-AdS$_4$ black hole by calculating the scrambling time $\tau_*$ and find that the leading term is logarithmic in entropy $\tau_*\sim\log S$ (see eq. (\ref{scramblingtime})). We use holographic entanglement entropy calculations to calculate the mutual information $I(A;B)$ and obtain both scrambling time $\tau_*$ and the Lyapunov exponent $\lambda_L$ by perturbing the black hole with rotating and charged shock waves. This work generalizes the chaotic behavior of black holes to the rotating black holes in the EMDA theory. Both $\tau_*$ and $\lambda_L$ depend on the new parameters, the dilaton and axion charges. \\
\indent From Figure \ref{fig:Lyapunovscaling}, we show that the instantaneous minimal Lyapunov exponent, for most of the cases, is bounded by $\kappa=2\pi T_H/(1-\mu\mathcal{L})$, and approaches $\kappa$ when $\frac{r_-}{r_+}\rightarrow1$, i.e. when the black hole approaches extremality. Furthermore, for larger values of AdS length $l$, the Lyapunov exponent also approaches $\kappa$ and saturates the bound. Only in the case where $l=1$ that we found the Lyapunov exponent exceeds $\kappa$, for some large values of $\mathcal{L}$. This violation may be caused by the existence of charges, especially the dilaton and axion charges. The fact that the Lyapunov exponent is approximately equal to $\kappa$ also supports the fast scrambling conjecture. For nonrotating black holes, the Lyapunov exponent should saturate the MSS bound \cite{Maldacena2016}, which is given by the black hole temperature. We see that from earlier works on rotating black holes, the upper bound for the Lyapunov index is modified by a factor of $\frac{1}{1-\mu\mathcal{L}}$, which can be greater than $2\pi T_H$ \cite{halder2019global,Malvimat2022BTZ,Malvimat2022KerrAdS}. \\
\indent We suspect that the cause for the violation of the $\kappa$ bound here is due to the existence of the dilaton or axion charges and in this work, we show such evidence using the plot given by Figure \ref{fig:Lyapunovscaling}. We plan to do the analytical calculations of the correction to the Lyapunov bound with the existence of dilaton and axion charges in future works. The fact that such a violation does not occur in the standard Kerr-AdS$_4$ black hole \cite{Malvimat2022KerrAdS} might help us to pinpoint the origin of the violation and this carries significant physical interests. In future works, we also plan to show that such a violation does not occur in a Kerr-NUT-AdS black hole \cite{Sakti2020KerrNUT} at similar conditions (for a small value of $l$). Furthermore, an investigation of the butterfly velocity and entanglement velocity due to localized shock waves \cite{Jahnke2018} in the Kerr-Sen-AdS$_4$ background might also help us to gain insight into the violation of the bound \footnote{We thank V. Jahnke for pointing this out and bringing the reference about the butterfly velocity of a static anisotropic black brane.}.  We plan to investigate it in future works as well.\\
\indent We also calculate the scrambling time and the Lyapunov exponent for the ultra-spinning version of the dyonic Kerr-Sen-AdS$_4$, i.e. the solution where the rotation parameter has its maximal value, $a\rightarrow l$. The ultra-spinning solution provides us with new functions that cannot be obtained by simply taking the limit of $a\rightarrow l$ from the standard dyonic Kerr-Sen-AdS$_4$. Due to the logarithmic behavior of the scrambling time, the ultra-spinning version also admits fast scrambling. However, the ratio between $\kappa$ and the Lyapunov exponent is quite large. For example, we have $\kappa/\lambda_L=\mathcal{C}\sim 12$ for the case which violates the RII with $\mathcal{R}=0.995016$ and $\mathcal{C}\sim2.5$ for the case with $\mathcal{R}=1$ (see Figure \ref{fig:plotultraspin}). From Figure \ref{fig:lambdaultraspinning}, we see that even though $l$ is small, the violation of the chaos bound does not present in the ultra-spinning case. We conclude that the ultra-spinning counterpart of the dyonic Kerr-Sen-AdS$_4$ black hole also admits chaotic behavior but the Lyapunov exponent is relatively smaller than the standard black hole, with larger $\mathcal{C}$.\\
\indent Due to the existence of charges, both in the black hole and the shock waves, we also have the scrambling time delay $\tau_d$.  The subleading term of $\tau_*$ in eq. (\ref{scramblingtimenormal}) and eq. (\ref{scramblingtimeultra}) represent the difference between the scrambling time of neutral black holes with the charged ones, and is denoted by $\Delta\tau_*$. For a perturbation that obeys $\delta S>0$, the aforementioned term is positive and hence prolongs the scrambling process. However, since the shock wave bounces inside the horizon at $r_b$ (that is assumed to be very close to the horizon), the role of the subleading term is to delay the onset of scrambling, following the analysis of \cite{horowitz_bouncing_2022}. We show that the scrambling process of a dyonic Kerr-Sen-AdS$_4$ black hole is also delayed by calculating the time that is needed to send a signal from the right asymptotic $\tau_d$ for the signal to meet the shock waves inside the black hole interior. The difference between $\tau_d$ and $\tau_0$, i.e. the time when the shock wave is sent from the left asymptotic, is equal to $\Delta\tau_*$ up to some corrections in the order of $\beta$. For the bounce to happen near the horizon, $\Delta\tau_*$ is much larger than the thermal time $\beta$. The scrambling delay time depends on the charges of the shock waves, $\mathcal{Q}$ and $\mathcal{P}$. This is the only function that depends on $\mathcal{Q},\mathcal{P}$, indicating that the role of the charges is to delay the scrambling. If we take $\mathcal{L}\rightarrow0$ (and also $\mathcal{P}\rightarrow0$), we recover the scrambling delay found in the non-rotating Reissner-Nordstr\"om black hole \cite{horowitz_bouncing_2022}.\\
\indent In conclusion, our calculations support the fast scrambling conjecture for the dyonic Kerr-Sen-AdS$_4$ black hole to some extent. However, since our calculations only focus on the non-extremal and near-extremal limits of the black hole, it is also important to understand the chaotic behavior of the extremal black hole. It is known that an extremal black hole and a non-extremal black hole are two distinct objects, and taking the extremal limit of a non-extremal black hole does not obtain the properties of the actual extremal black hole \cite{Carroll2009,Prihadi_2023}. Therefore, the chaotic behavior of extremal dyonic Kerr-Sen-AdS$_4$ black hole deserves more investigation for future works.
\subsection*{Appendix: Stress tensor from null-shell formalism}
In this Appendix, we derive the stress tensor of the rotating charged shock waves using the null-shell formalism in \cite{Poisson2002} and hence we adopt their notation here. Starting with metric in eq. (\ref{metricUV}), we construct the normal vector to the shocks following constant $u=r_*-\tau$ path, $k^\mu=g^{\mu\nu}\partial_\nu u$, which is given by
\begin{equation}
k^\mu\partial_\mu=\frac{1}{F}\bigg(\frac{\Delta}{\tilde{f}}\partial_r+\partial_\tau-h_\tau\partial_z\bigg),
\end{equation}
while the other tangent generator is given by $e_z^\mu=\delta_z^\mu$. From here, we can obtain the transverse null vector by solving
\begin{equation}
N_\mu N^\mu=0,\;\;\; N_\mu k^\mu=-1,\;\;\;N_\mu e^\mu_z=0,
\end{equation}
which results in
\begin{equation}
N_\mu dx^\mu=\frac{F}{2}\bigg(d\tau+\frac{\tilde{f}}{\Delta}dr\bigg).
\end{equation}
The transverse extrinsic curvature can then be calculated from
\begin{equation}
C_{AB}=\nabla_\nu N_\mu e^\mu_A e^\nu_B=\frac{\Delta h'}{2\tilde{f}}\delta_A^z\delta_B^z\equiv\frac{\sigma}{2}\delta_A^z\delta_B^z,
\end{equation}
with $h'(r)=dh(r)/dr$.
The difference between $C_{AB}$ evaluated in the perturbed manifold with the unperturbed one is proportional to the stress-energy tensor, as given by eq. (\ref{stresstensor}). For vanishing all black hole parameters except those corresponding to the Reissner-Nordstr\"om black hole, the result recovers the stress tensor in \cite{horowitz_bouncing_2022}.
\section*{Acknowledgement}
FPZ would like to thank Kemenristek, the Ministry of Research, Technology, and Higher Education, Republic of Indonesia for financial support. HLP would like to thank GTA Institut Teknologi Bandung for financial support. HLP would like to thank the members of the Theoretical Physics Groups of Institut Teknologi Bandung for their hospitality. S. A. was
supported by an appointment to the Young Scientist Training
Program at the Asia Pacific Center for Theoretical Physics
(APCTP) through the Science and Technology Promotion
Fund and Lottery Fund of the Korean Government. This was
also supported by the Korean Local Governments -
Gyeongsangbuk-do Province and Pohang City.

\bibliography{EEBHSRT.bib}

\end{document}